%% file: main.tex
\documentclass[conference]{IEEEtran}
\IEEEoverridecommandlockouts

\usepackage{soul}
\usepackage{graphicx}
\usepackage{booktabs}
\usepackage{float}
\usepackage{overpic}
\usepackage{float}  
\usepackage{subfloat}
\usepackage{stfloats} 
\usepackage[skip=0.5\baselineskip]{caption}
\usepackage{bm}
\usepackage{amsmath}
\usepackage{booktabs}
\usepackage{multirow}
\usepackage{multicol}
\usepackage{enumitem}
\usepackage{subcaption}
\usepackage{threeparttable}
\usepackage{xspace}
\usepackage{color}
\usepackage[normalem]{ulem}
\usepackage{algorithmicx,algorithm}
\usepackage{algpseudocode}
\usepackage{algorithm,algpseudocode}
\usepackage{marvosym}   
\usepackage{bm}
\usepackage{makecell}
\usepackage{url}
\usepackage{tcolorbox}  
\tcbuselibrary{breakable}   
\usepackage{wasysym}

\usepackage{xspace}
\newcommand{\etal}{\emph{et al.}\xspace}
\newcommand{\eg}{\emph{e.g.,}\xspace}
\newcommand{\ie}{\emph{i.e.,}\xspace}

\usepackage{cite}
\usepackage{textcomp}
\usepackage{amsfonts}
\def\BibTeX{{\rm B\kern-.05em{\sc i\kern-.025em b}\kern-.08em
    T\kern-.1667em\lower.7ex\hbox{E}\kern-.125emX}}
\begin{document}
\begin{sloppypar}

\makeatletter
\newcommand{\linebreakand}{%
  \end{@IEEEauthorhalign}
  \hfill\mbox{}\par
  \mbox{}\hfill\begin{@IEEEauthorhalign}
}
\makeatother

\title{Large Language Model Distilling Medication Recommendation Model
}

\author{\IEEEauthorblockN{Qidong Liu}
\IEEEauthorblockA{\textit{Xi'an Jiaotong University} \\
\textit{City University of Hong Kong}\\
Xi'an, China \\
liuqidong@stu.xjtu.edu.cn}
\and
\IEEEauthorblockN{Xian Wu \Letter}
\IEEEauthorblockA{\textit{Jarvis Research Center, Tencent YouTu Lab}\\
Shenzhen, China \\
kevinxwu@tencent.com}
\thanks{\Letter \ \text{Corresponding Authors}}
\and
\IEEEauthorblockN{Xiangyu Zhao \Letter}
\IEEEauthorblockA{\textit{City University of Hong Kong}\\
Hong Kong, China \\
xianzhao@cityu.edu.hk}
\linebreakand
\IEEEauthorblockN{Yuanshao Zhu}
\IEEEauthorblockA{\textit{Southern University of Science and Technology} \\
\textit{City University of Hong Kong}\\
Shenzhen, China \\
zhuys2019@mail.sustech.edu.cn}
\\
\IEEEauthorblockN{Feng Tian \Letter}
\IEEEauthorblockA{\textit{Xia'an Jiaotong University}\\
Xi'an, China \\
fengtian@mail.xjtu.edu.cn}
\and
\IEEEauthorblockN{Zijian Zhang}
\IEEEauthorblockA{\textit{Jilin University} \\
\textit{City University of Hong Kong}\\
Changchun, China \\
zhangzj2114@mails.jlu.edu.cn}
\\
\IEEEauthorblockN{Yefeng Zheng}
\IEEEauthorblockA{\textit{Jarvis Research Center, Tencent YouTu Lab}\\
Shenzhen, China \\
yefengzheng@tencent.com}
}

\maketitle

\begin{abstract}
    The recommendation of medication is a vital aspect of intelligent healthcare systems, as it involves prescribing the most suitable drugs based on a patient's specific health needs. Unfortunately, many sophisticated models currently in use tend to overlook the nuanced semantics of medical data, while only relying heavily on identities. Furthermore, these models face significant challenges in handling cases involving patients who are visiting the hospital for the first time, as they lack prior prescription histories to draw upon. To tackle these issues, we harness the powerful semantic comprehension and input-agnostic characteristics of Large Language Models (LLMs). Our research aims to transform existing medication recommendation methodologies using LLMs. In this paper, we introduce a novel approach called \textbf{L}arg\textbf{E} langu\textbf{A}ge mo\textbf{D}el distilling m\textbf{E}dication \textbf{R}ecommendation (LEADER). We begin by creating appropriate prompt templates that enable LLMs to suggest medications effectively. However, the straightforward integration of LLMs into recommender systems leads to an out-of-corpus issue specific to drugs. We handle it by adapting the LLMs with a novel output layer and a refined tuning loss function. Although LLM-based models exhibit remarkable capabilities, they are plagued by high computational costs during inference, which is impractical for the healthcare sector. To mitigate this, we have developed a feature-level knowledge distillation technique, which transfers the LLM's proficiency to a more compact model. Extensive experiments conducted on two real-world datasets, MIMIC-III and MIMIC-IV, demonstrate that our proposed model not only delivers effective results but also is efficient.
    To ease the reproducibility of our experiments, we release the implementation code online~\footnote{https://github.com/liuqidong07/LEADER-pytorch}.
\end{abstract}


\input{1Introduction}

\input{2Preliminary}
\input{3Method}

\input{4Experiment}

\input{5RelatedWork}

\input{6Conclusion}


\bibliographystyle{IEEEtran}
\bibliography{main}

\appendix

\input{7Appendix}
\end{sloppypar}
\end{document}

%% file: 1Introduction.tex
\section{Introduction}  \label{sec:intro}

\begin{table}[!t]
\centering
\caption{The investigation of current medication recommendation models. ``\CIRCLE'' means the type of input necessary or ability for inference. ``\Circle'' means no such type of input or inability for inference. ``\RIGHTcircle'' means the type of input alternative.}
\resizebox{0.5\textwidth}{!}{
\begin{tabular}{c|ccc|cc}
\toprule[1pt]
\multirow{2}{*}{\textbf{Model}} & \multicolumn{3}{c|}{\textbf{Input}} & \multicolumn{2}{c}{\textbf{Inference}} \\ \cmidrule{2-6} 
& \textbf{Diagnosis} & \textbf{Procedure} & \textbf{Medication} & \textbf{Single-visit} & \textbf{Multi-visit} \\ 
\midrule
\textbf{RETAIN}~\cite{choi2016retain} & \CIRCLE & \CIRCLE & \Circle & \CIRCLE & \CIRCLE \\
\textbf{G-Bert}~\cite{shang2019pre} & \CIRCLE & \Circle & \CIRCLE & \Circle & \CIRCLE \\
\textbf{GAMENet}~\cite{shang2019gamenet} & \CIRCLE & \CIRCLE & \RIGHTcircle & \CIRCLE & \CIRCLE \\
\textbf{SafeDrug}~\cite{yang2021safedrug} & \CIRCLE & \CIRCLE & \Circle & \CIRCLE & \CIRCLE \\
\textbf{MICRON}~\cite{yang2021change} & \CIRCLE & \CIRCLE & \CIRCLE & \Circle & \CIRCLE \\
\textbf{COGNet}~\cite{wu2022conditional} & \CIRCLE & \CIRCLE & \CIRCLE & \Circle & \CIRCLE \\
\textbf{REFINE}~\cite{bhoi2023refine} & \CIRCLE & \CIRCLE & \CIRCLE & \Circle & \CIRCLE \\ 
\midrule
\textbf{LEADER (Ours)} & \CIRCLE & \CIRCLE & \CIRCLE & \CIRCLE & \CIRCLE \\ 
\bottomrule[1pt]
\end{tabular}
}
\label{tab:preliminary}
\end{table}

Prescription, as a crucial aspect of patient treatment, is labor-intensive and requires specialized expertise~\cite{rahmawati2020physician}. Automated medication recommender systems offer potential relief for overburdened healthcare professionals, by providing decision support~\cite{ali2023deep}. 
Contemporary medication recommendation models primarily focus on generating drug recommendations based on patients' diagnostic and procedural data. While significant advancements have been achieved, two primary challenges persist: 
(i) \textbf{Lack of Semantic Understanding}: Existing models~\cite{choi2016retain,shang2019gamenet,yang2021safedrug,shang2019pre} predominantly capture the collaborative information among medications, diagnoses, and procedures by their identity data. However, the importance of semantic understanding, especially in medical contexts~\cite{li2023chatdoctor}, is frequently overlooked in medication recommendation. 
(ii) \textbf{Challenges with Single-Visit Patients}: Prescription history is a critical factor in current prescription practices, as indicated by recent studies~\cite{bhoi2023refine,wu2022conditional,yang2021change}. As shown in Table~\ref{tab:preliminary}, models like MICRON~\cite{yang2021change}, COGNet~\cite{wu2022conditional} and REFINE~\cite{bhoi2023refine} incorporate historical medication records as their input for enhanced performance. 
However, this reliance on historical data poses a significant challenge in recommending for first-time hospital visitors, termed \textit{single-visit patients}. Excluding single-visit patients is unacceptable in real-world healthcare systems, indicating a crucial area for improvement.


The advent of large language models~\cite{zhao2023survey} presents an opportunity to enhance existing medication recommender systems. 
On the one hand, extensive studies have confirmed the robust semantic understanding capabilities of large language models~\cite{wei2022chain}. 
This enables the refinement of medication recommendations from a medical semantics perspective. 
On the other hand, LLMs process natural language as input, making them inherently agnostic to the types and number of input variables~\cite{borisov2022language}. 
Consequently, unlike some existing medication recommendation models, LLM-based medication recommenders can incorporate any conceivable variables, including patients' profiles and historical prescriptions, into the model. 
This flexibility allows them to cater to all patients, irrespective of whether a patient has a documented medical history. 
Addressing the two challenges mentioned before, the application of large language models to the medication recommendation task emerges as a compelling and attractive solution.

Several pioneering works~\cite{chen2023large, wu2023survey} have taken the initial steps to integrate large language models with recommender systems. However, their direct application to the medication recommendation task is hindered by two significant problems:
(i) \textbf{Out-of-corpus Problem}. 
Numerous studies~\cite{yang2023large, lin2023multi, zheng2023generative} have explored the creation of input prompts to engage LLMs. 
Nevertheless, the incompatibility between the natural language output and the required in-corpus drugs persists. 
This challenge may result in recommendations from LLM-based recommender systems that are not part of the drug set, potentially compromising recommendation performance. 
For instance, an LLM might generate a medication name that cannot be verified in the drug bank, leading to a failed recommendation.
(ii) \textbf{High Inference Cost Problem}. 
LLMs often suffer from high inference latency and memory issues~\cite{zhou2023opportunities}, given their billions of parameters. 
While general applications can leverage cloud computing to meet real-time requirements for LLM-based services, the deployment of medical services within healthcare institutions, such as hospitals, is common due to privacy concerns~\cite{gruendner2019ketos}. 
Besides, equipping each medical center with a high-performance computing platform poses a logistical challenge. 
Therefore, a more efficient solution for LLM-based medication recommendation is imperative.

To address the aforementioned challenges, we introduce the \textbf{L}arg\textbf{E} L\textbf{A}nguage Mo\textbf{D}el \textbf{E}nhanced Medication \textbf{R}ecommendation by Distillation (\textbf{LEADER}). 
In our approach, to adapting LLMs for medication recommendation, we first develop appropriate prompt templates to activate the LLM's semantic understanding ability. 
Specifically, for the out-of-corpus issue, we enhance the LLM by introducing a new output layer with a corresponding training loss. 
Following supervised fine-tuning, the LLM gains the capability for medication recommendation and exhibits exceptional performance.
However, the application of the LLM-based model is hindered by high inference costs. To address this issue, we delve into transferring the formidable capabilities of the LLM to a small model. 
In detail, a feature-level distillation method is devised to augment the small medication recommendation model based on the adapted LLM.
The contributions of this paper are as follows:
\begin{itemize}[leftmargin=*]
    \item We validate the robust capability of LLMs for the medication recommendation task through the modification of the output layer and fine-tuning loss specific to LLMs. To the best of our knowledge, we are the first to explore the integration of medication recommendation and large language models.
    \item We introduce a feature-level knowledge distillation method to enhance the small model using LLMs, resulting in a highly efficient and effective medication recommendation model.
    \item Extensive experiments are conducted on two public datasets, namely MIMIC-III and MIMIC-IV. The experimental results consistently demonstrate that the proposed LEADER model outperforms current baselines.
\end{itemize}

%% file: 2Preliminary.tex
\begin{table}[]
\centering
\caption{Notions used in LEADER. ``med.'', ``diag.'' and ``proc.'' are the abbreviations of medication, diagnosis and procedure.}
\resizebox{\linewidth}{!}{%
\begin{tabular}{ll}
\toprule[1pt]
\textbf{Notation} & \textbf{Description} \\ 
\midrule
$\mathcal{X}^{(z)}$  &  The EHR records of the patient $z$      \\
$\mathcal{P}^{(z)}$ & The lingual prompt of patient's EHR record \\
$P$ & The profile features of one patient \\
$T_z$  &  The number of hospital visits for patient $z$     \\  
\midrule
$\mathcal{M}_i$, $\mathcal{D}_i$, $\mathcal{P}_i$    &   The set of med., diag., and proc. codes  \\ 
$\mathbf{E}_m$, $\mathbf{E}_d$, $\mathbf{E}_p$  &  Embedding matrices for med., diag. and proc. codes       \\
$\mathcal{E}_{Diag}$, $\mathcal{E}_{Proc}$, $\mathcal{E}_{Med}$ & The encoder for med., diag. and proc. sets \\
$\mathcal{E}_{p}$ & The profile feature encoder \\
$\mathcal{E}_{Visit}$ & The medical record visit encoder \\
$\mathbf{h}$ & The hidden state from last transformer layer of LLM \\
$\mathbf{W}_{CLS}$ & The classification output layer for LLM \\
$\mathbf{W}_{proj}$ & The linear projection layer for distillation \\
\midrule
$\hat{\mathbf{y}}$  &  The predicted probability for each med. \\
$\mathbf{y}$   &   Labels of recommended med.  \\
\bottomrule[1pt]
\end{tabular}
\label{tab:pre_notation}}
\end{table}

\section{Preliminary}   \label{sec:preliminary}

Electronic Health Records (EHR) is one essence of an intelligent healthcare system, which collects patients' detailed and procedural medical data. In EHR, the patient's data can be handled by their hospital visits. Assume, there are $N$ patients in the database, then the records of the patient $z$ are represented as $\mathcal{X}^{(z)}=[\mathcal{X}^{(z)}_1,...,\mathcal{X}^{(z)}_i,...,\mathcal{X}^{(z)}_{T_z}]$, where $T_z$ is visit number of this patient. For simplicity, the patient stamp $(z)$ is omitted in the following. Since diagnosis and procedures are vital for prescription in the real world~\cite{shang2019gamenet,yang2021safedrug}, these two elements with medications are included in each visit record. In the visit $i$, the record is denoted as $\mathcal{X}_i=\{\mathcal{M}_i, \mathcal{D}_i, \mathcal{P}_i\}$. A patient may take several drugs and get multiple diagnoses and procedures, so let $\mathcal{M}_i=\{m_1,...,m_j,...,m_{|\mathcal{M}|}\}$, $\mathcal{D}_i=\{d_1,...,d_j,...,d_{|\mathcal{D}|}\}$, $\mathcal{P}_i=\{p_1,...,p_j,...,p_{|\mathcal{P}|}\}$ denote the set of medication, diagnosis and procedure, respectively. $|\mathcal{M}|$, $|\mathcal{D}|$ and $|\mathcal{P}|$ represent the totals of them. Some demographic characteristics of patients, such as age, gender, etc., are also vital, which are marked as $P$. We list out the important notations of this paper in Table~\ref{tab:pre_notation}.

Medication recommendation aims to give out the proper medication set $\mathcal{M}_T$ given all possible medical data of this patient. As mentioned before, many existing methods adopt the patient's historical prescriptions for a more accurate recommendation, which requires the patient to have multiple visits, \ie $T>1$. However, they cannot handle the single-visit patient with $T=1$. In this paper, we explore to derive the model for both types of patients. Therefore, we define the problem respectively. For \textbf{single-visit} patients, recommend $\mathcal{M}_T$ given $\{\mathcal{D}_T, \mathcal{P}_T, P\}$. For  \textbf{multi-visit} patients, give out $\mathcal{M}_T$ based on $\{\mathcal{X}_1,...,\mathcal{X}_{T-1}, \mathcal{D}_T, \mathcal{P}_T, P\}$.

%% file: 3Method.tex
\section{Method}

In this section, the details of the proposed LEADER are introduced. At first, we will present the overview in Section~\ref{sec:method_overview}. Then, the modification of the LLM for medication recommendation is illustrated in Section~\ref{sec:method_LLM}. In Section~\ref{sec:method_distill}, we will illustrate the distillation method for transferring the powerful semantic understanding ability of the LLM to a small model. At last, the procedures of optimization and inference are detailed in Section~\ref{sec:method_opt}.

\begin{figure*}[t]
\centering
\includegraphics[width=1\linewidth]{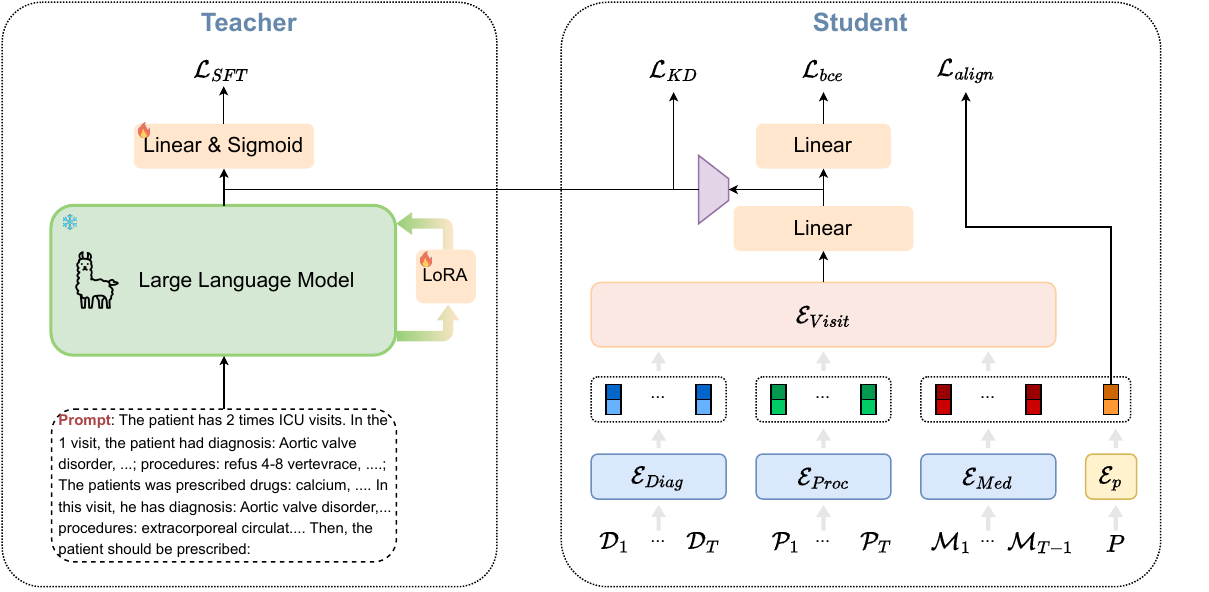}
\caption{The framework overview of the proposed LEADER, which consists of two training stages. The first stage is to supervised fine-tune the \textit{Teacher} medication recommendation model, \ie large language model. In the second stage, we train the designed \textit{Student} medication recommendation by knowledge distillation. For high efficiency, only the student model is used for inference.}
\label{fig:framework}
\end{figure*}

\subsection{Overview}   \label{sec:method_overview}

    The overview of the proposed LEADER is shown in Figure~\ref{fig:framework}. To utilize the LLM, we design the prompt template to format the electronic health record of the patient into natural language. Then, the output and fine-tuning loss function are modified to better fit the medication recommendation task, which can be considered as a multi-class classification problem. Though the LLMs have been proven to have brilliant ability~\cite{zeng2022glm,touvron2023llama,openai2023gpt4} recently, they face the problem of high inference cost, which is hardly accepted by the healthcare system. Thus, we explore transferring the powerful ability from LLM to the designed small model by the proposed knowledge distillation method. In the diagram, the LLM-based model and small model are represented as ``Teacher'' and ``Student'', respectively. We will train the student model from scratch with the ground-truth label and knowledge distillation loss from the well-fine-tuned teacher model.

\subsection{LLM for Medication Recommendation} \label{sec:method_LLM}

    The input and output of the large language model are both natural languages, while they are non-semantic identities in conventional medication recommendation models~\cite{wu2022conditional,yang2021safedrug,shang2019gamenet}, such as ``Medication ID: 2''.
    Thus, to apply the LLMs to medication recommendation, we have to fill such a gap. On the one hand, we design the proper \textbf{prompt templates} to format the electronic health records into natural language, which can be input to LLMs directly. On the other hand, lingual output for recommendation by LLMs faces the out-of-corpus challenge~\cite{bao2023bi,lin2023multi}, so we substitute the original language head with a classification \textbf{output layer}. 
    Correspondingly, the objective for fine-tuning LLMs is modified. Next, we will detail the prompt templates and output layer in the following parts.

    \subsubsection{\textbf{Prompt Templates}}
    We design the prompt template $\mathcal{T}$ to derive the lingual representation $\mathcal{P}^{(z)}$ of the patient's EHR, which can instruct the LLM to understand the health condition of the patient. The devised template is as follows:

    \begin{tcolorbox}[
            colframe=gray,
            width=1\linewidth,
            arc=1mm, 
            auto outer arc,
            title={Input Prompt Template},
            breakable,]
            
            The patient has \underline{$<$VISIT\_NUM$>$} times ICU visits. 
            
            \textcolor{blue}{In the 1 visit, the patient had diagnosis: \ul{$<$DIAG\_NAME$>$}, ..., \ul{$<$DIAG\_NAME$>$}; procedures: \ul{$<$PROC\_NAME$>$}, ..., \ul{$<$PROC\_NAME$>$}; The patient was prescribed drugs: \ul{$<$MED\_NAME$>$}, ..., \ul{$<$MED\_NAME$>$}. In the 2 visit, .... }
            
            In this visit, the patient has diagnosis: \underline{$<$DIAG\_NAME$>$}, ..., \underline{$<$DIAG\_NAME$>$}; procedures: \underline{$<$PROC\_NAME$>$}, ..., \underline{$<$PROC\_NAME$>$}. Then, the patient should be prescribed:
            
    \end{tcolorbox}

    In the template, the places underlined will be filled in with EHR data. ``$<$VISIT\_NUM$>$'' is the number of hospital visits for one patient. The part in blue represents the historical records $\{\mathcal{X}_1,...,\mathcal{X}_{T-1}\}$ of the patient. However, we have argued that the patients who first visit the hospital are either important. For these single-visit patients, they do not have this part in the prompt. Besides, the diagnosis, procedure and medication are represented by their name to utilize the semantic understanding ability of LLMs. Thus, ``$<$DIAG\_NAME$>$'', ``$<$PROC\_NAME$>$'' and ``$<$MED\_NAME$>$'' are all standard medical terms in the prompt. After the prompt construction, the LLMs can conduct an understanding for medication recommendation from the lingual input.

    \subsubsection{\textbf{Output Layer}}
    Most existing LLM-based recommender systems~\cite{geng2022recommendation,bao2023tallrec} output the name or identity of the recommendations in natural language but face the out-of-corpus challenge. To tackle this problem, we substitute the pre-trained word token generation layer with a linear layer accompanied by a sigmoid. Then, the outputs from the modified LLM are the probability of every medication.
    \begin{equation}    \label{eq:llm_out}
        \hat{\mathbf{y}} = \sigma(\mathbf{W}_{CLS} \cdot \mathbf{h})
    \end{equation}
    \noindent where $\hat{\mathbf{y}} \in \mathbb{R}^{|\mathcal{M}| \times 1}$ and $\mathbf{h} \in \mathbb{R}^{d_h \times 1}$ are the predicted probability of medication and the hidden states from the last transformer layer in LLMs. $\mathbf{W}_{CLS} \in \mathbb{R}^{|\mathcal{M}| \times d_h}$ is a learnable weight matrix and $\sigma(\cdot)$ represents the sigmoid function. For the final recommendation, a threshold $\gamma$ will be set. When $y_k > \gamma$, the medication $k$ will be included in the prescribed medication set.

    \subsubsection{\textbf{Optimization}}
    Since we renew the output layer of the LLM, supervised fine-tuning (SFT) is necessary. At the same time, SFT can benefit the LLMs to complete specific tasks~\cite {li2023chatdoctor,wang2023huatuo}. However, the conditional language modeling objective~\cite{zeng2022glm,touvron2023llama} is unsuitable for the modified LLM, because the output layer is for classification. To better fit the medication recommendation task and the output layer, we modify the loss function of SFT as follows:
    \begin{equation}    \label{eq:sft}
        \mathcal{L}_{SFT} = - \sum_{i=1}^N \mathbf{y}^{(i)}log(\hat{\mathbf{y}}^{(i)}) + (1 - \mathbf{y}^{(i)})log(1 - \hat{\mathbf{y}}^{(i)})
    \end{equation}
    \noindent In the equation, $\mathbf{y}$ is the ground-truth medication labels. It is worth noting that fine-tuning all parameters of the LLM is extremely costly. Therefore, we adopt LoRA~\cite{hu2021lora} fine-tuning in this paper, which only updates sets of low-rank matrices while freezing the pre-tained weights of the LLM. Let $\{\mathbf{A}_i, \mathbf{B}_i\}_{i=1}^L$ denotes the sets of trainable matrices, where $L$ is the number of layers accompanied by LoRA layer. Then, during the SFT, only the parameters $\mathbf{W}_{CLS}$ and $\{\mathbf{A}_i, \mathbf{B}_i\}_{i=1}^L$ are trainable and initialized by normal distribution.

\subsection{Enhancement by Distillation} \label{sec:method_distill}

    Though the LLMs possess powerful semantic understanding abilities, they require high inference memory and latency. It is unacceptable to the healthcare system, so we aim to transfer the abilities of LLMs to a relatively small model. The knowledge distillation~\cite{gou2021knowledge} is a promising way, but the student model architecture and specific distillation method still need to be addressed.
    
    \subsubsection{\textbf{Student Model Design}}
    Considering the efficiency issue, the identities, instead of the semantic terms, are adopted in the student model. As mentioned in Section~\ref{sec:preliminary}, the input variables can be written as $\{\mathcal{D}_1,...,\mathcal{D}_T;\mathcal{P}_1,...,\mathcal{P}_T;\mathcal{M}_1,...,\mathcal{M}_{T-1};P\}$, where $\mathcal{D}$, $\mathcal{P}$ and $\mathcal{M}$ are sets of diagnosis, procedure and medications. 

    To capture collaborative information from each type of set, we design three homogeneous encoders for them, denoted as $\mathcal{E}_{Diag}$, $\mathcal{E}_{Proc}$ and $\mathcal{E}_{Med}$, respectively. For brevity, we only take the $\mathcal{E}_{Diag}$ for illustration. We first derive an embedding table $\mathbf{E}_d \in \mathbb{R}^{|\mathcal{D}| \times d_e}$, where each row refers to the unique code of diagnosis. $d_e$ represents the dimension of the embedding table. Then, the set of diagnosis codes $\mathcal{D}_i$ are transformed into a set of vectors by $\mathbf{E}_d$, denoted as $\bar{\mathcal{D}}_i = [\mathbf{d}_1,..., \mathbf{d}_{|\mathcal{D}_i|}]$. 
    Next, we propose to adopt a transformer architecture to encode the inter-relationship contained in each set. The pair of multi-head attention and feed-forward networks consist of one transformer layer, which can be written as:
    \begin{equation}
        \mathbf{M}={\rm LayerNorm}(\bar{\mathcal{D}}_i, {\rm MultiHead}(\bar{\mathcal{D}}_i,\bar{\mathcal{D}}_i,\bar{\mathcal{D}}_i))
    \end{equation}
    \noindent where ${\rm LayerNorm(\cdot)}$ and ${\rm MultiHead(\cdot)}$ represents the layer normalization and multi-head attention, respectively. The other component of the transformer layer is the feed-forward network accompanied by a residual connection, which can be formulated as follows:
    \begin{equation}
        \hat{\mathcal{D}}^{(1)}={\rm LayerNorm}(\mathbf{M}, {\rm FNN}(\textbf{M}))
    \end{equation}
    \noindent where ${\rm FNN}(\cdot)$ is one trainable linear layer. The output of the first transformer layer is denoted as $\hat{\mathcal{D}}^{(1)}$, which is a sequence of vectors. Then, we impose the average pooling to the output from the last transformer layer of $\mathcal{E}_{Diag}$ and get the representation of the diagnosis set, \ie $\mathbf{D}_i \in \mathbb{R}^{d_t}$.
    \begin{equation}
        \mathbf{D}_i = {\rm Avg\_pool}(\hat{\mathcal{D}}^{(L_d)})
    \end{equation}
    \noindent where $L_d$ denotes the number of transformer layer in $\mathcal{E}_{Diag}$. By the diagnosis encoder, the input diagnosis records $\{\mathcal{D}_1,...,\mathcal{D}_T\}$ are converted to a set of vectors, \ie $[\mathbf{D}_1,...,\mathbf{D}_T]$. Similarly, we can get the representation of procedure and medication sets by $\mathcal{E}_{Proc}$ and $\mathcal{E}_{Med}$ with the same structure as $\mathcal{E}_{Diag}$.

    Then, we devise a visit encoder $\mathcal{E}_{Visit}$ to capture the historical health conditions of the patients. In specific, $\mathcal{E}_{Visit}$ is also stacked by several transformer layers, which is the same as $\mathcal{E}_{Diag}$. Thus, $\mathcal{E}_{Visit}$ will encode the sequence of diagnosis records into one embedding $\tilde{\mathbf{D}}$, which can be written as follows:
    \begin{equation}
        \tilde{\mathbf{D}} = \mathcal{E}_{Visit}([\mathbf{D}_1,...,\mathbf{D}_T])
    \end{equation}
    In the same way, we can get the representation of historical procedure and medication records, denoted as $\tilde{\mathbf{P}}$ and $\tilde{\mathbf{M}}$. It is worth noting that the three types of records share the visit encoder $\mathcal{E}_{Visit}$, because such a design can not only shrink the number of parameters but also help learn the shared medical knowledge~\cite{shang2019pre}.
    
    Another challenge for the student model is the difficulty for single-visit patients because the input of medication records to $\mathcal{E}_{Visit}$ is empty when $T=1$. Here, we propose using the profile information as a pseudo medication record since the profile can reflect the patient's health condition. In detail, the profile feature, such as age, is discretized and then encoded by embedding matrices. All representations of profile features are concatenated and then projected to an $d_t$ dimensional vector, marked as $\mathbf{P}$. The profile vector will be inserted into the sequence of medication records, so the medication input to $\mathcal{E}_{Visit}$ are changed to $[\mathbf{M}_1,...,\mathbf{M}_{T-1},\mathbf{P}]$.

    Finally, we concatenate the $\tilde{\mathbf{D}}$, $\tilde{\mathbf{P}}$ and $\tilde{\mathbf{M}}$, and adopt two linear layers for final medication recommendation.
    \begin{equation}
        \hat{\mathbf{y}} = \sigma(\mathbf{W}_2 (\mathbf{W}_1 \cdot [\tilde{\mathbf{D}}||\tilde{\mathbf{P}}||\tilde{\mathbf{M}}] + \mathbf{b}_1)+\mathbf{b}_2)
    \end{equation}
    \noindent where $\mathbf{W}_1 \in \mathbb{R}^{3d_t \times d_t}$, $\mathbf{W}_2 \in \mathbb{R}^{d_t \times |\mathcal{M}|}$, $\mathbf{b}_2 \in \mathbb{R}^{1 \times d_t}$ and $\mathbf{b}_2 \in \mathbb{R}^{1 \times |\mathcal{M}|}$ are trainable parameters. Then, the loss function for the ground-truth label is written as:
    \begin{equation}    \label{eq:bce}
        \mathcal{L}_{bce} = - \sum_{i=1}^N \mathbf{y}^{(i)}log(\hat{\mathbf{y}}^{(i)}) + (1 - \mathbf{y}^{(i)})log(1 - \hat{\mathbf{y}}^{(i)})
    \end{equation}

    \subsubsection{\textbf{Knowledge Distillation}}
    In order to transfer the powerful ability of the LLM-based model to the student model, we propose a feature-level knowledge distillation. Since LLMs are skilled in memorizing~\cite{tirumala2022memorization,biderman2023emergent}, they can predict the samples in the training set with relatively high accuracy. This will cause the prediction of the training set from LLMs to be similar to the ground-truth label, which is not suitable for distillation. Therefore, we propose to distill the student model by the hidden state from LLMs. 

    The hidden state $\mathbf{h}$ is the representation from the last transformer layer of LLMs. In the conventional pre-trained LLMs, this hidden state is used to generate the word token via a linear layer, so it contains comprehensive semantic information. In the modified LLMs, since $\mathbf{h}$ can output the probabilities of medications accompanied by a classification layer, it is also suitable to guide the student model considering the task similarity. 

    However, the representation in the student model is still in a different space of $\mathbf{h}$, because there is no semantic input to the student model. Therefore, we design a trainable projector to transform the hidden state into the representation space of LLM. Then, the loss for knowledge distillation can be written as:
    \begin{equation}    \label{eq:kd}
        \mathcal{L}_{KD} = \frac{1}{N}\sum^N_{i=1} \mid \mathbf{h}_i - \mathbf{W}_{proj} \cdot (\mathbf{W}_1 \cdot [\tilde{\mathbf{D}}_i||\tilde{\mathbf{P}}_i||\tilde{\mathbf{M}}_i] + \mathbf{b}_1) \mid^2
    \end{equation}
    \noindent where $\mathbf{W}_{proj} \in \mathbb{R}^{d_t \times d_h}$ is weight of projection layer. Note that all the parameters of the student model and $\mathbf{W}_{proj}$ are updated during the distillation, while the parameters of LLM are frozen.

    \subsubsection{\textbf{Profile Alignment}} \label{sec:method_align}
    Due to the design of the profile features as a pseudo medication record, our model can recommend for single-visit patients. However, the representations of the profile and medication set are actually in different spaces, which causes difficulty in training. As a result, to align the two different types of representations, we design a profile alignment method.

    Inspired by the contrastive learning for modality alignment in multimodal research~\cite{li2022blip,radford2021learning}, we propose a contrastive loss to align profile and medication sets. For better performance~\cite{chen2020simple}, we first project the representation of profile $\mathbf{P}$ and the target medication set $\mathbf{M}_T$ to a new space:
    \begin{equation}
        \begin{aligned}
            \mathbf{Z}_P &= \mathbf{W}_{proj}^P \cdot \mathbf{P} \\
            \mathbf{Z}_M &= \mathbf{W}_{proj}^M \cdot \mathbf{M}_T
        \end{aligned}
    \end{equation}
    \noindent where $\mathbf{W}_{proj}^P \in \mathbb{R}^{d_t \times d_t}$ and $\mathbf{W}_{proj}^M \in \mathbb{R}^{d_t \times d_t}$ are the projection matrices. Let $[\mathbf{Z}_P^1,...,\mathbf{Z}_P^B]$ and $[\mathbf{Z}_M^1,...,\mathbf{Z}_M^B]$ denote one batch of profile and medication representations, where $B$ is the batch size. We consider $\mathbf{Z}_P^i$ and $\mathbf{Z}_M^j$ as a positive pair, when $i=j$. Then, the contrastive loss for the profile can be defined as:
    \begin{equation}
        \mathcal{L}_{PM}=-\frac{1}{B} \sum_{i=1}^{B} \log \frac{\exp(sim(\mathbf{Z}_P^i, \mathbf{Z}_M^i) / \tau)}{\sum_{j=1}^B \mathbb{I}_{[i \neq j]} \exp(sim(\mathbf{Z}_P^i, \mathbf{Z}_M^j) / \tau)}
    \end{equation}
    \noindent where $\mathbb{I}_{[i \neq j]}$ represents an indicator function. $\tau$ denotes the temperature parameter in the loss. In the same way, we can also derive the contrastive loss $\mathcal{L}_{MP}$ for medication. Thus, the alignment loss is the sum of these two losses:
    \begin{equation}    \label{eq:align}
        \mathcal{L}_{align} = \sum_{N} \mathcal{L}_{PM} + \mathcal{L}_{MP}
    \end{equation}

\let\oldnl\nl
\newcommand{\nonl}{\renewcommand{\nl}{\let\nl\oldnl}}
\begin{algorithm}[t]
\caption{Train and Inference Process of LEADER} \label{alg:train}
\raggedright

\begin{algorithmic} [1]
    \State Indicate the prompt template $\mathcal{T}$.
    \State Construct the lingual input $\mathcal{P}$ according to $\mathcal{X}$.
    \State Indicate the hyper-parameters $\alpha$, $\beta$ and $\tau$. 
\end{algorithmic}

\textbf{Train Stage 1} 
\setcounter{algorithm}{2}
\begin{algorithmic} [1]
    \makeatletter
    \setcounter{ALG@line}{3}
    \State Substitute the language generation head with the classification head. Freeze all the pre-trained parameters $\Theta_{LLM}$ of the LLM.
    \For {a batch of samples $B_p$ in $\mathcal{P}$}
        \State Input $B_p$ to the LLM and get the hidden state $\mathbf{h}$.
        \State Output the probability of each medication by Equation~\eqref{eq:llm_out}.
        \State Fine-tune $\mathbf{W}_{CLS}$ and $\{\mathbf{A}_i, \mathbf{B}_i\}_{i=1}^L$ by Equation~\eqref{eq:sft}.
    \EndFor
    \State Get the teacher model LEADER(T).
\end{algorithmic}

\textbf{Train Stage 2}
\setcounter{algorithm}{9}
\begin{algorithmic} [1]
    \makeatletter
    \setcounter{ALG@line}{10}
    \For {a batch of samples $B_p$, $B_x$ in $\mathcal{P}$, $\mathcal{X}$}
        \State Input $B_x$ to student model and get the BCE loss by Equation~\eqref{eq:bce}.
        \State Input $B_p$ to LEADER(T) to get the hidden state $\mathbf{h}$, and calculate the loss for distillation by Equation~\eqref{eq:kd}.
        \State Update the parameters of the student model $\Theta_{stu}$ and $\mathbf{W}_{proj}$ by the Equation~\eqref{eq:all}.
    \EndFor
    \State Get the student model LEADER(S).
\end{algorithmic}

\textbf{Inference}
\setcounter{algorithm}{15}
\begin{algorithmic} [1]
    \makeatletter
    \setcounter{ALG@line}{16}
    \State Transform the patient's EHR $\mathcal{X}^{(z)}$ to $\mathcal{P}^{(z)}$ by $\mathcal{T}$.
    \State Input $\mathcal{X}^{(z)}$ to LEADER(S) and get the recommendation.
    \State Input $\mathcal{P}^{(z)}$ to LEADER(T) and get the recommendation.
\end{algorithmic}
\end{algorithm}

\subsection{Train and Inference} \label{sec:method_opt}

    The proposed LEADER needs two-stage optimization. In the first stage, we need to optimize the modified LLM by Equation~\eqref{eq:sft}. The fine-tuned LLM will act as the teacher model, dubbed LEADER(T). In the second stage, the student model denoted as LEADER(S), is trained from scratch by the combination of loss from the ground-truth label, knowledge distillation and profile alignment, \ie
    \begin{equation}    \label{eq:all}
        \mathcal{L} = \mathcal{L}_{bce}+\alpha \cdot \mathcal{L}_{KD} + \beta \cdot \mathcal{L}_{align}
    \end{equation}
    \noindent where $\alpha$ and $\beta$ are the hyper-parameters to adjust the scale of distillation and alignment. After the optimization, both LEADER(T) and LEADER(S) can complete the medication recommendation task, but have distinct input formats. To show the process of training and inference more clearly, we conclude the Algorithm~\ref{alg:train}.

    Firstly, we indicate some necessary hyper-parameters and construct the natural language input for LLM (line 1-3). Then, at the first stage, the modified LLM is supervised fine-tuned by the derived lingual dataset (line 4-10). The fine-tuned modified LLM can be used for both distillation or medication recommendation directly. At the second training stage, the EHR formatted in natural language and identity are absorbed by the teacher and student model, respectively (line 11-13). Then, we update the student model by the combination of BCE, distillation and alignment loss (line 14-16). In terms of inference, we can either adopt the LEADER(S) or LEADER(T) for the final recommendation (line 17-19).

%% file: 4Experiment.tex
\section{Experiment}

In this section, we will analyze the proposed LEADER by comprehensive experiments on two real-world datasets. We explore the following Research Questions (\textbf{RQ}) to illustrate the findings:

\begin{itemize}[leftmargin=*]
    \item \textbf{RQ1}: How the proposed LEADER perform compared with current state-of-the-art medication recommendation models and LLM-based recommendation models?
    \item \textbf{RQ2}: Do all designs for LEADER take effect?
    \item \textbf{RQ3}: How do the designed knowledge distillation and profile alignment affect the performance of LEADER?
    \item \textbf{RQ4}: Can the proposed student model conduct medication recommendation with a high efficiency?
\end{itemize}

\subsection{Experimental Settings}

\subsubsection{\textbf{Dataset}}

    The datasets used in the experiments are from Medical Information Mart for Intensive Care (MIMIC)~\footnote{https://mimic.mit.edu/}. There are two versions available currently, \ie MIMIC-III and MIMIC-IV. MIMIC-III collects data from 2001 to 2012, while MIMIC-IV contains records from 2008 to 2019. 
    We follow the preprocessing of the previous works~\cite{shang2019gamenet,yang2021safedrug}. Due to the space limitation, we leave the more detailed introduction of the datasets to \textbf{Appendix~\ref{sec:appendix_dataset}}.

\subsubsection{\textbf{Baselines}}
    In the experiments, we compare our LEADER with several state-of-the-art \textbf{Medication Recommendation Models} (RETAIN~\cite{choi2016retain}, G-Bert~\cite{shang2019pre}, GAMENet~\cite{shang2019gamenet}, SafeDrug~\cite{yang2021safedrug}, MICRON~\cite{yang2021change}, COGNet~\cite{wu2022conditional} REFINE~\cite{bhoi2023refine}) and \textbf{LLM-based Recommendation Models} (TALLRec~\cite{bao2023tallrec}, BIGRec~\cite{bao2023bi}, E4SRec~\cite{li2023e4srec}). The detailed introduction and implementation of the baselines can be seen in \textbf{Appendix~\ref{sec:appendix_baseline}}.
    We compare the modified LLM proposed in Section~\ref{sec:method_LLM}, denoted as \textbf{LEADER(T)}. Also, the distilled student model is marked as \textbf{LEADER(S)} in the following experiments.

\begin{table*}[!t]
\centering
\tabcolsep=0.05cm   
\caption{The overall results of competing baselines and LEADER on MIMIC-III. The boldface refers to the highest score and the underline indicates the best result of the models. ``\textbf{{\Large *}}'' indicates the statistically significant improvements (\ie two-sided t-test with $p<0.05$) over the best baseline. ``-'' represents the model cannot acquire the corresponding results due to the inability to the single-visit patients or TALLRec has no PRAUC due to its output of medication name instead of probability}.
\label{tab:exp_mimic3}
\resizebox{1\textwidth}{!}{
\centering
\begin{tabular}{c|ccc|ccc|ccc}
\toprule[1pt]
\multirow{2}{*}{\textbf{Model}} & \multicolumn{3}{c|}{\textbf{Overall}} & \multicolumn{3}{c|}{\textbf{Multi-visit}} & \multicolumn{3}{c}{\textbf{Single-visit}} \\ \cmidrule{2-10} 
 & \textbf{PRAUC} & \textbf{Jaccard} & \textbf{F1} & \textbf{PRAUC} & \textbf{Jaccard} & \textbf{F1} & \textbf{PRAUC} & \textbf{Jaccard} & \textbf{F1} \\ \midrule
RETAIN & 0.7513 $\pm$ 0.0025 & 0.4943 $\pm$ 0.0023 & 0.6516 $\pm$ 0.0022 & 0.7580 $\pm$ 0.0020 & 0.5106 $\pm$ 0.0023 & 0.6674 $\pm$ 0.0022 & 0.7337 $\pm$ 0.0067 & 0.4811 $\pm$ 0.0053 & 0.6403 $\pm$ 0.0049 \\
G-Bert & - & - & - & 0.6904 $\pm$ 0.0017 & 0.4578 $\pm$ 0.0019 & 0.6186 $\pm$ 0.0018 & - & - & - \\
GAMENet & 0.7605 $\pm$ 0.0011 & 0.5024 $\pm$ 0.0010 & 0.6595 $\pm$ 0.0008 & 0.7638 $\pm$ 0.0023 & 0.5070 $\pm$ 0.0028 & 0.6635 $\pm$ 0.0025 & 0.7451 $\pm$ 0.0053 & 0.4840 $\pm$ 0.0038 & 0.6442 $\pm$ 0.0036 \\
SafeDrug & 0.7582 $\pm$ 0.0020 & 0.5054 $\pm$ 0.0024 & 0.6621 $\pm$ 0.0021 & 0.7623 $\pm$ 0.0029 & 0.5095 $\pm$ 0.0027 & 0.6658 $\pm$ 0.0024 & 0.7416 $\pm$ 0.0044 & 0.4900 $\pm$ 0.0043 & 0.6481 $\pm$ 0.0042 \\
MICRON & - & - & - & 0.7651 $\pm$ 0.0027 & 0.5110 $\pm$ 0.0025 & 0.6741 $\pm$ 0.0023 & - & - & - \\
COGNet & - & - & - & 0.7771 $\pm$ 0.0019 & \underline{0.5275 $\pm$ 0.0021} & \underline{0.6805 $\pm$ 0.0019} & - & - & - \\
REFINE & - & - & - & 0.7791 $\pm$ 0.0017 & 0.5235 $\pm$ 0.0018 & 0.6794 $\pm$ 0.0017 & - & - & - \\ 
\midrule
TALLRec & - & 0.4420 $\pm$ 0.0021 & 0.6053 $\pm$ 0.0021 & - & 0.4476 $\pm$ 0.0015 & 0.6110 $\pm$ 0.0015 & - & 0.4205 $\pm$ 0.0055 & 0.5839 $\pm$ 0.0057 \\
BIGRec & 0.7577 $\pm$ 0.0019 & 0.4946 $\pm$ 0.0020 & 0.6525 $\pm$ 0.0018 & 0.7591 $\pm$ 0.0017 & 0.4949 $\pm$ 0.0022 & 0.6528 $\pm$ 0.0024 & 0.7521 $\pm$ 0.0054 & 0.4931 $\pm$ 0.0046 & 0.6509 $\pm$ 0.0056 \\
E4SRec & 0.7663 $\pm$ 0.0021 & 0.5062 $\pm$ 0.0019 & 0.6627 $\pm$ 0.0024 & 0.7717 $\pm$ 0.0016 & 0.5123 $\pm$ 0.0023 & 0.6686 $\pm$ 0.0020 & 0.7447 $\pm$ 0.0041 & 0.4819 $\pm$ 0.0049 & 0.6393 $\pm$ 0.0052 \\
\midrule
\textbf{LEADER(T)} & \textbf{0.7816 $\pm$ 0.0015}* & \textbf{0.5391 $\pm$ 0.0015}* & \textbf{0.6921 $\pm$ 0.0014}* & \textbf{0.7854 $\pm$ 0.0015}* & \textbf{0.5450 $\pm$ 0.0021}* & \textbf{0.6971 $\pm$ 0.0018}* & \underline{0.7590 $\pm$ 0.0046}* & \textbf{0.5090 $\pm$ 0.0044}* & \textbf{0.6668 $\pm$ 0.0041}* \\ 
\textbf{LEADER(S)} & \underline{0.7795 $\pm$ 0.0025}* & \underline{0.5175 $\pm$ 0.0022}* & \underline{0.6737 $\pm$ 0.0019}* & \underline{0.7830 $\pm$ 0.0019} & 0.5208 $\pm$ 0.0020 & 0.6768 $\pm$ 0.0017 & \textbf{0.7631 $\pm$ 0.0056}* & \underline{0.5038 $\pm$ 0.0062}* & \underline{0.6614 $\pm$ 0.0057}* \\ 
\bottomrule[1pt]
\end{tabular}
}
\end{table*}

\begin{table*}[!t]
\centering
\tabcolsep=0.05cm   
\caption{The overall results of competing baselines and LEADER on MIMIC-IV. The boldface refers to the highest score and the underline indicates the best result of the models. ``\textbf{{\Large *}}'' indicates the statistically significant improvements (\ie two-sided t-test with $p<0.05$) over the best baseline. ``-'' represents the model cannot acquire the corresponding results due to the inability to the single-visit patients or TALLRec has no PRAUC due to its output of medication name instead of probability}.
\label{tab:exp_mimic4}
\resizebox{1\textwidth}{!}{
\begin{tabular}{c|ccc|ccc|ccc}
\toprule[1pt]
\multirow{2}{*}{\textbf{Model}} & \multicolumn{3}{c|}{\textbf{Overall}} & \multicolumn{3}{c|}{\textbf{Multi-visit}} & \multicolumn{3}{c}{\textbf{Single-visit}} \\ \cmidrule{2-10} 
 & \textbf{PRAUC} & \textbf{Jaccard} & \textbf{F1} & \textbf{PRAUC} & \textbf{Jaccard} & \textbf{F1} & \textbf{PRAUC} & \textbf{Jaccard} & \textbf{F1} \\ \midrule
RETAIN & 0.6574 $\pm$ 0.0055 & 0.4152 $\pm$ 0.0044 & 0.5688 $\pm$ 0.0043 & 0.6576 $\pm$ 0.0044 & 0.4161 $\pm$ 0.0038 & 0.5693 $\pm$ 0.0040 & 0.6588 $\pm$ 0.0055 & 0.4165 $\pm$ 0.0035 & 0.5707 $\pm$ 0.0042 \\
G-Bert & - & - & - & 0.6237 $\pm$ 0.0028 & 0.3727 $\pm$ 0.0021 & 0.5169 $\pm$ 0.0022 & - & - & - \\
GAMENet & 0.6720 $\pm$ 0.0030 & 0.4336 $\pm$ 0.0032 & 0.5871 $\pm$ 0.0030 & 0.6731 $\pm$ 0.0030 & 0.4339 $\pm$ 0.0020 & 0.5877 $\pm$ 0.0021 & 0.6671 $\pm$ 0.0049 & 0.4292 $\pm$ 0.0041 & 0.5819 $\pm$ 0.0040 \\
SafeDrug & 0.6706 $\pm$ 0.0025 & 0.4295 $\pm$ 0.0027 & 0.5820 $\pm$ 0.0024 & 0.6752 $\pm$ 0.0031 & 0.4331 $\pm$ 0.0018 & 0.5860 $\pm$ 0.0017 & 0.6641 $\pm$ 0.0072 & 0.4214 $\pm$ 0.0073 & 0.5749 $\pm$ 0.0073 \\
MICRON & - & - & - & 0.6660 $\pm$ 0.0041 & 0.4414 $\pm$ 0.0027 & 0.5951 $\pm$ 0.0027 & - & - & - \\
COGNet & - & - & - & 0.6873 $\pm$ 0.0034 & \underline{0.4638 $\pm$ 0.0028} & \underline{0.6119 $\pm$ 0.0026} & - & - & - \\
REFINE & - & - & - & 0.6977 $\pm$ 0.0042 & 0.4538 $\pm$ 0.0047 & 0.6063 $\pm$ 0.0044 & - & - & - \\ 
\midrule
TALLRec & - & 0.4190 $\pm$ 0.0021 & 0.5731 $\pm$ 0.0022 & - & 0.4301 $\pm$ 0.0033 & 0.5841 $\pm$ 0.0030 & - & 0.3988 $\pm$ 0.0045 & 0.5527 $\pm$ 0.0043 \\
BIGRec & 0.6756 $\pm$ 0.0029 & 0.4357 $\pm$ 0.0022 & 0.5887 $\pm$ 0.0025 & 0.6792 $\pm$ 0.0029 & 0.4385 $\pm$ 0.0035 & 0.5969 $\pm$ 0.0027 & 0.6702 $\pm$ 0.0044 & 0.4301 $\pm$ 0.0032 & 0.5773 $\pm$ 0.0028 \\
E4SRec & 0.6823 $\pm$ 0.0021 & 0.4396 $\pm$ 0.0024 & 0.5905 $\pm$ 0.0020 & 0.6845 $\pm$ 0.0030 & 0.4438 $\pm$ 0.0039 & 0.6008 $\pm$ 0.0030 & 0.6773 $\pm$ 0.0052 & 0.4324 $\pm$ 0.0049 & 0.5798 $\pm$ 0.0033 \\
\midrule
\textbf{LEADER(T)} & \textbf{0.7120 $\pm$ 0.0024}* & \textbf{0.4779 $\pm$ 0.0021}* & \textbf{0.6296 $\pm$ 0.0020}* & \textbf{0.7238 $\pm$ 0.0031}* & \textbf{0.4895 $\pm$ 0.0033}* & \textbf{0.6400 $\pm$ 0.0032}* & \underline{0.6881 $\pm$ 0.0039}* & \textbf{0.4539 $\pm$ 0.0026}* & \textbf{0.6071 $\pm$ 0.0026}* \\ 
\textbf{LEADER(S)} & \underline{0.7020 $\pm$ 0.0022}* & \underline{0.4483 $\pm$ 0.0025}* & \underline{0.6005 $\pm$ 0.0026}* & \underline{0.6994 $\pm$ 0.0037} & 0.4500 $\pm$ 0.0031 & 0.6023 $\pm$ 0.0029 & \textbf{0.7033 $\pm$ 0.0041}* & \underline{0.4420 $\pm$ 0.0039}* & \underline{0.5946 $\pm$ 0.0039}* \\ 
\bottomrule[1pt]
\end{tabular}
}
\vspace{-4mm}
\end{table*}

\subsubsection{\textbf{Implementation Details}}

    All experiments in this paper are conducted on the Intel Xeon Gold 6133 platform with Tesla V100 32G GPUs. The code is based on Python 3.9.5 and PyTorch 1.12.0. As for the LLM-based medication recommendation, \ie LEADER(T), and all LLM-based recommendation baselines, we adopt the LLaMA-7B~\footnote{https://github.com/facebookresearch/llama/tree/llama\_v1}~\cite{touvron2023llama} as the foundation model in this paper. Besides, we adopt LoRA~\cite{hu2021lora} as the fine-tuning method for all LLM-based models.
    Due to the space limitation, we leave more implementation details to \textbf{Appendix~\ref{sec:appendix_implement}}.
    To facilitate the reproduction of our model, we release the code online~\footnote{https://anonymous.4open.science/r/LEADER-447E}.

\subsubsection{\textbf{Evaluation Metrics}}

    Following the previous works~\cite{shang2019gamenet,yang2021safedrug,wu2022conditional,bhoi2023refine}, we apply three common metrics to evaluate the proposed model, \ie \textit{Precision-Recall AUC} (\textbf{PRAUC} $\uparrow$), \textit{Jaccard Similarity Score} (\textbf{Jaccard} $\uparrow$) and \textit{Average F1 Score} (\textbf{F1} $\uparrow$). To guarantee the robustness of the experimental results, we adopt bootstrapping sampling during the test process. In detail, we randomly sample $80\%$ samples in each round and the metrics shown below are the averaged on $10$-round tests.

\subsection{Overall Performance (RQ1)}

    To respond to the research question (\textbf{RQ1}), we reveal the performance comparison between the proposed method and competitors in Table~\ref{tab:exp_mimic3} and Table~\ref{tab:exp_mimic4}. Then, we address the analysis of the results.

    Overall, LEADER(T) performs a strong lead compared with all of the other models on two datasets, which indicates the semantic understanding ability of the LLM. At the same time, the distilled student model, marked as LEADER(S), also outperforms the medication recommendation and LLM-based models. This phenomenon shows the success of the designed distillation enhancement. 

    Then, we probe the performance comparison according to different patient groups. As mentioned before, some recent baselines, \eg G-Bert, MICRON, COGNet and REFINE, consider the historical medication records as one of the necessary inputs, so they do not have the results for single-visit patients. We first observe the multi-visit patient group. G-Bert performs the worst, because it does not take patient's procedures into consideration. Then, we can find that the three baselines (MICRON, COGNet and REFINE), which model the historical prescriptions explicitly, can outperform the other competitors in the multi-visit group. Such comparison illustrates utilizing previous drug records can actually benefit the recommendation for the current visit. The proposed LEADER(T) can surpass all models consistently due to the powerful ability of the LLM. As for the designed LEADER(S), it can outperform others on the PRAUC metric, but is worse than COGNet on Jaccard and F1. We think the reason lies in that COGNet adopts beam search to generate the final recommendations, but it faces efficiency issues.

    In terms of the performance in single-visit and overall groups, GAMENet and SafeDrug are better than RETAIN, because they model the relations between medications more carefully by EHR graph and molecule graph. However, they still underperform the two variants of the proposed LEADER consistently. On the one hand, LEADER can utilize the historical information and surpass baselines in the multi-visit group largely, which contributes to the overall scores. On the other hand, due to the semantic understanding ability of LLM, LEADER(T) and LEADER(S) both surpass competitors in the single-visit group on two datasets. It is worth noting that the distilled LEADER(S) is even better than LEADER(T) in the single-visit group under the PRAUC metric. This phenomenon indicates the benefits of the combination of collaborative signals from the student model and semantic information from LLM. 

    As for the LLM-based models, TALLRec even underperforms some medication recommendation models. The inferior performance is caused by the direct output of medication name, highlighting the \textbf{out-of-corpus} problem. 
    BIGRec and E4SRec can get higher recommending accuracy on both overall and single-visit groups, which indicates a powerful semantic understanding ability of the LLM. However, they still lag behind the proposed LEADER. For BIGRec, the reason lies in the sub-optimal grounding method. In terms of E4SRec, it only integrates the collaborative signals into LLM, resulting in underutilization of the LLM.

    From the analysis, we conclude that the proposed LLM-based medication recommendation model shows a greater \textbf{semantic understanding ability} and \textbf{single-visit ability} than conventional models. Besides, the designed distillation method can actually enhance the derived student model.

\begin{table*}[!t]
\centering
\caption{The ablation study on two datasets. Due to limited space, only PRAUC scores are shown in the table.}
\label{tab:exp_ablation}
\resizebox{0.9\textwidth}{!}{
\begin{tabular}{c|ccc|ccc}
\toprule[1pt]
\multirow{2}{*}{\textbf{Model}} & \multicolumn{3}{c|}{\textbf{MIMIC-III}} & \multicolumn{3}{c}{\textbf{MIMIC-IV}}  \\ \cmidrule{2-7} 
 & \textbf{Overall} & \textbf{Multi-visit} & \textbf{Single-visit} & \textbf{Overall} & \textbf{Multi-visit} & \textbf{Single-visit} \\ \midrule
\textbf{LEADER(S)} & 0.7795 $\pm$ 0.0025 & 0.7830 $\pm$ 0.0019 & 0.7631 $\pm$ 0.0056 & 0.7020 $\pm$ 0.0022 & 0.6994 $\pm$ 0.0037 & 0.7033 $\pm$ 0.0041 \\ \midrule
\textit{w/o} KD & 0.7673 $\pm$ 0.0026 & 0.7720 $\pm$ 0.0034 & 0.7464 $\pm$ 0.0052 & 0.6840 $\pm$ 0.0031 & 0.6853 $\pm$ 0.0044 & 0.6768 $\pm$ 0.0062 \\
\textit{w/o} feature-KD & 0.7672 $\pm$ 0.0024 & 0.7730 $\pm$ 0.0034 & 0.7448 $\pm$ 0.0058 & 0.6846 $\pm$ 0.0023 & 0.6865 $\pm$ 0.0028 & 0.6792 $\pm$ 0.0047 \\
\textit{w/o} align & 0.7774 $\pm$ 0.0017 & 0.7836 $\pm$ 0.0031 & 0.7579 $\pm$ 0.0054 & 0.6967 $\pm$ 0.0026 & 0.6987 $\pm$ 0.0039 & 0.6939 $\pm$ 0.0026  \\
\textit{w/o} shared $\mathcal{E}_{Visit}$ & 0.7781 $\pm$ 0.0010 & 0.7830 $\pm$ 0.0022 & 0.7613 $\pm$ 0.0050 & 0.6985 $\pm$ 0.0021 & 0.7001 $\pm$ 0.0035 & 0.6923 $\pm$ 0.0045  \\
\bottomrule[1pt]
\end{tabular}
}
\end{table*}

\subsection{Ablation Study (RQ2)}

    To verify the effectiveness of each proposed component for LEADER, we conduct ablation experiments. The results are shown in Table~\ref{tab:exp_ablation}. Firstly, we aim to validate how the designed feature-level knowledge distillation has an effect on the student model. \textit{w/o KD} represents we remove the KD loss directly during the training of LEADER(S), while \textit{w/o feature-KD} denotes using the KL-divergence of output probability between student and teacher model as the KD loss~\cite{hinton2015distilling}. From the results, we can find that these two variants both underperform the proposed LEADER(S) by a large margin. The drastic performance drop indicates that the feature-level knowledge distillation can actually enhance the collaborative student model. Also, the designed feature-level KD is more suitable for knowledge transfer from LLM than traditional output-level KD.

    Then, we seek to explore whether our design for the student model is reasonable. In Table~\ref{tab:exp_ablation}, \textit{w/o align} means that we leave out the profile alignment module proposed in Section~\ref{sec:method_align}. The experimental results illustrate that the alignment benefits the single-visit group more, which contributes to the overall performance elevation. The reason may be that the alignment can refine the representation of the profile, which is considered the only medication record for single-visit patients. \textit{w/o shared $\mathcal{E}_{Visit}$} represents that the designed student model adopts a split visit encoder for diagnosis, procedure and medication. This variant is worse than LEADER(S), which shows that the shared encoder can help learn more general medical knowledge. As the response to \textbf{RQ2}, we can conclude that the designed feature-level KD and other components in the student model are all beneficial to LEADER(S).
    Furthermore, to validate the effect of various LLM, such as QWen, we leave the related experimental results and analysis to \textbf{Appendix~\ref{sec:appendix_llm}}.

\begin{figure}[t]
\centering
\includegraphics[width=1\linewidth]{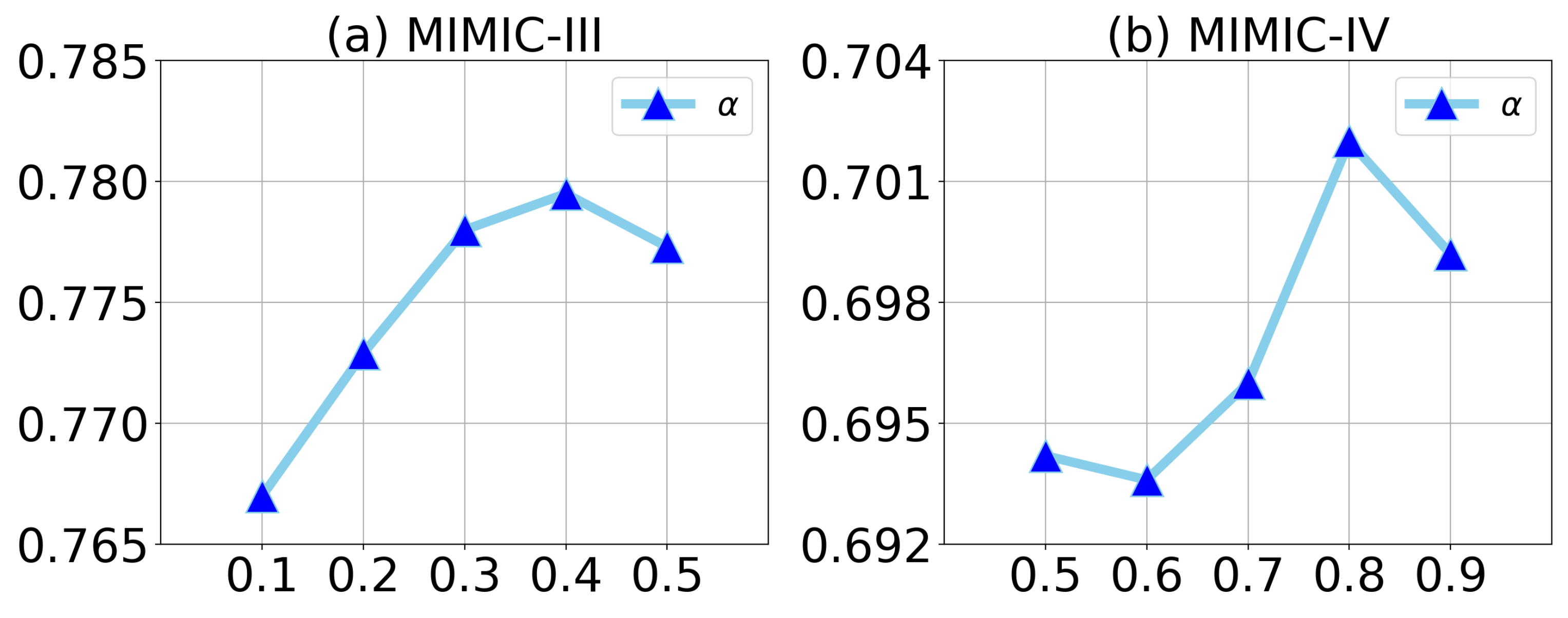}
\caption{The results of experiments for the weight of knowledge distillation loss $\alpha$ on two datasets.}
\label{fig:exp_KD}
\end{figure}

\begin{figure}[t]
\centering
\includegraphics[width=1\linewidth]{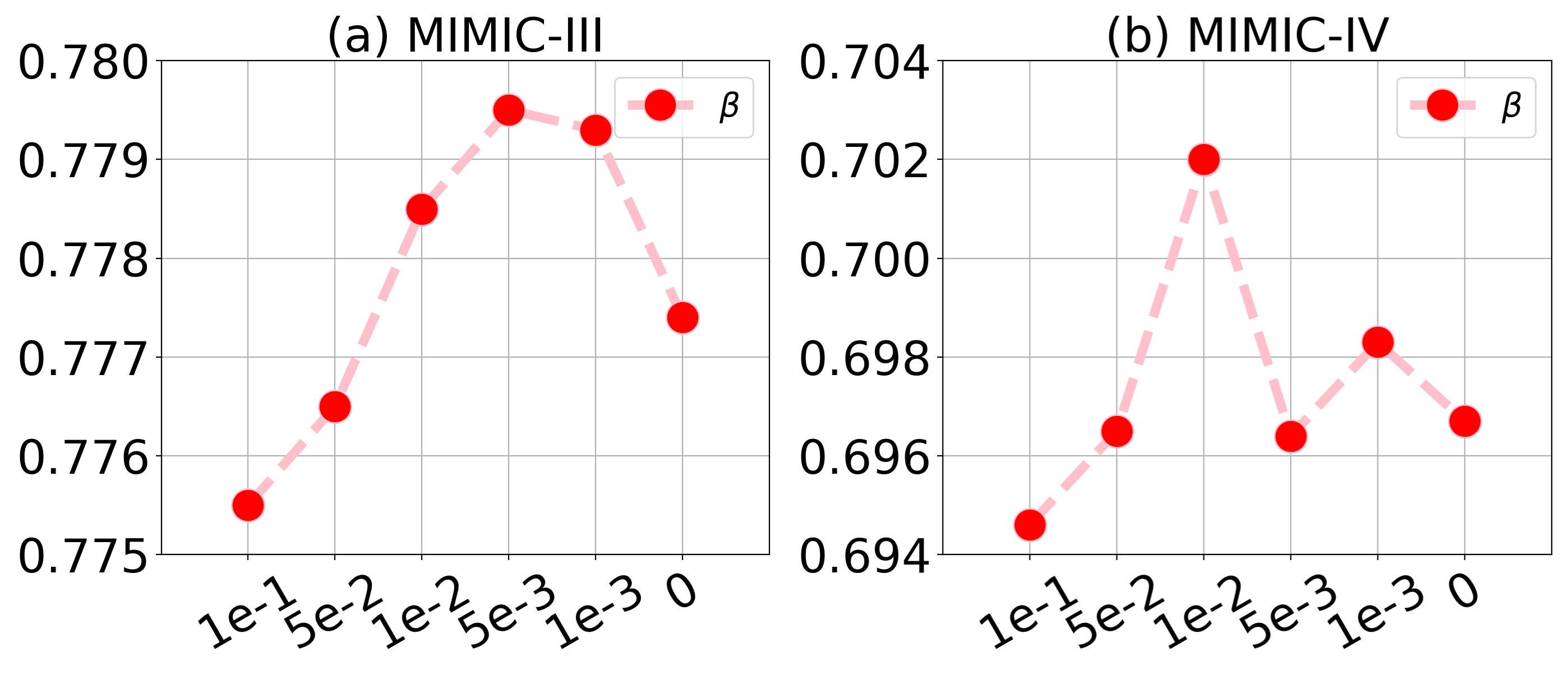}
\caption{The results of experiments for the weight of alignment loss $\beta$ on two datasets.}
\label{fig:exp_CL}
\end{figure}

\subsection{Hyper-parameter Analysis (RQ3)}

    To answer the \textbf{RQ3}, we adjust the strength of knowledge distillation and profile alignment during the training. The Figure~\ref{fig:exp_KD} and \ref{fig:exp_CL} shows the performance change according to $\alpha$ and $\beta$, respectively. We observe that the performance of LEADER(S) rises when $\alpha$ increases in a certain range. This phenomenon indicates that the knowledge transfer from the LLM-based teacher model can benefit the collaborative model. However, too large a scale of KD loss will confuse the model training towards the ground-truth labels, so the PRAUC score drops with the continual increase of $\alpha$. The best value of $\alpha$ for MIMIC-III is $0.4$. In terms of profile alignment, the figure shows the general performance trend is up at first and then down with $\beta$ change from $0.1$ to $0$. The reason why PRAUC increases at first is that too large a strength of contrastive loss will be harmful to the model convergence. In contrast, since the alignment can help refine the representation of the profile, the PRAUC drops when $\beta$ then decreases to $0$. As a result, the best $\beta$ for MIMIC-III is $5e^{-3}$.

\subsection{Efficiency Analysis (RQ4)}


\begin{figure}[t]
\centering
\includegraphics[width=1\linewidth]{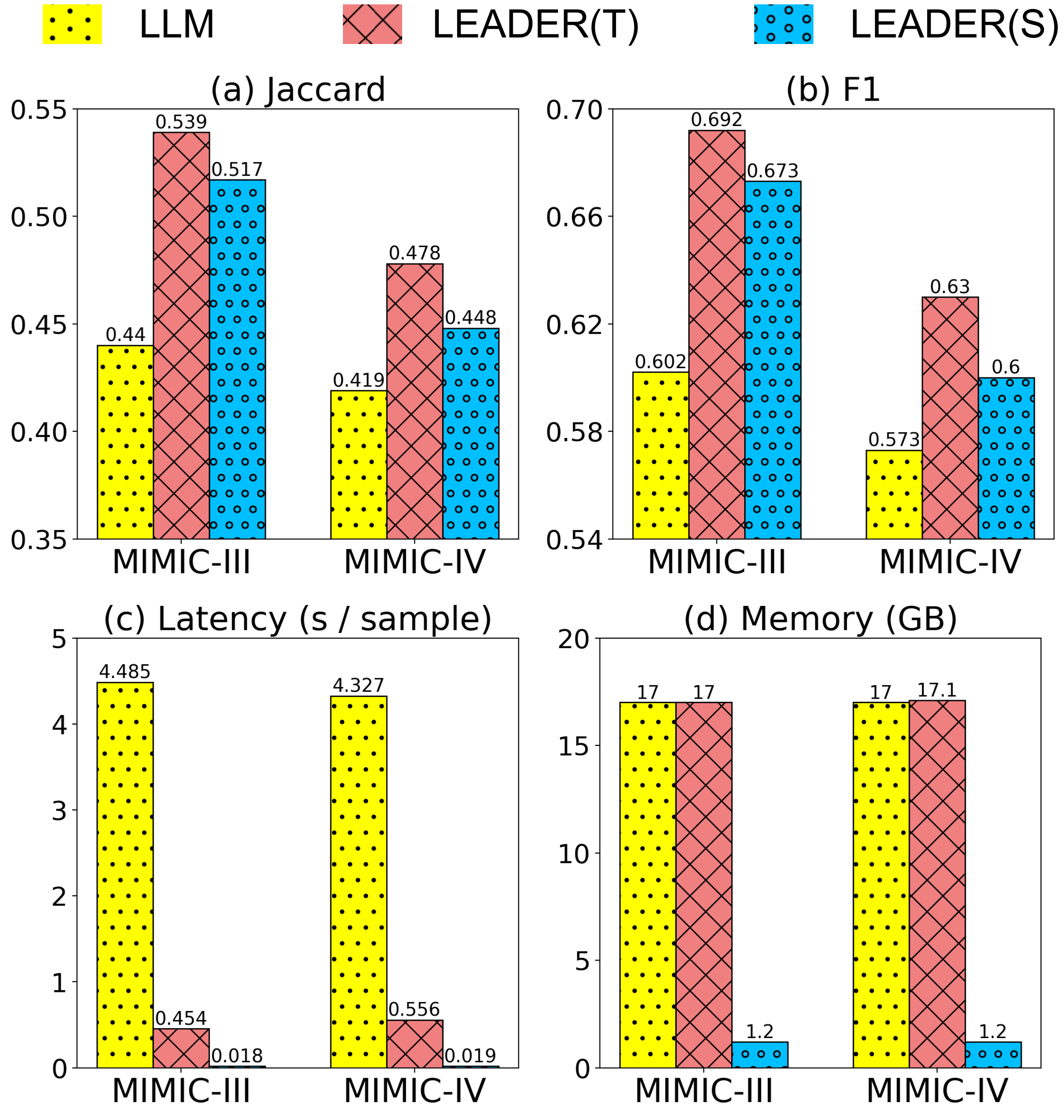}
\caption{The inference cost comparison between LLM and distilled small model, which is measured by (a) averaged inference time per sample and (b) necessary GPU memory.}
\label{fig:exp_efficiency}
\end{figure}

    As mentioned before, inference efficiency is an important issue in medical applications. Thus, we compare the efficiency between the LLM-based model and collaborative student model to respond \textbf{RQ4}. We apply the latency and GPU memory to measure the efficiency. In detail, the latency is calculated by averaging the total inference time of the test set on the number of test samples. Thus, the latency represents the average waiting time to complete the recommendation for one patient. The memory is the minimum GPU memory requirement for inference. As shown in Figure~\ref{fig:exp_efficiency}, 
    we can find that LEADER(T) has a shorter latency than general TALLRec. It is caused by the beam search during the word token generation, while the modified LLM can give out the probability in one run. In a word, the proposed modification of LLM can elevate effectiveness and efficiency simultaneously. However, both LLM-based medication recommendation models still pose the \textbf{high inference costs} problem. From the results, the proposed LEADER(S) can implement $25 \times \sim 30 \times$ inference acceleration and only requires about $1/15$ GPU memory compared with the LEADER(T). In summary, the designed LLM-distilled medication recommendation model can get a better trade-off between performance and efficiency.

%% file: 5RelatedWork.tex
\section{Related Works}


\subsection{Large Language Model for Recommendation}

Recently, the utilization of a large language model has been a hotspot in the recommender system community~\cite{wu2023survey,li2023large,fan2023recommender}. There are two main lines of work in the field of large language model for recommendation (LLM4Rec). One is tunable LLM4Rec, which often conducts fine-tuning to adapt the LLMs to the recommendation task better. P5~\cite{geng2022recommendation} firstly formulates the recommendation into a language generation task and then integrates various recommendation tasks into a unified language model. It fine-tunes a T5~\cite{raffel2020exploring} model to equip it with the ability to generate recommendations. Then, the applications of larger models, such as LlaMA and ChatGLM, bring more performance elevation. 
TALLRec~\cite{bao2023tallrec} designs proper instructions integrated with the user's historical records and fine-tunes a LlaMA-7b to complete the sequential recommendation. It is worth noting that parameter-efficient fine-tuning~\cite{hu2021lora,liu2023moelora} is often adopted because of efficiency issues. InstructRec~\cite{zhang2023recommendation} fabricates the preference, intention and task form to compose the prompt input. To further understand the users and shrink the prompt length, PALR~\cite{chen2023palr} inserts the summary of the user profile rather than raw features into the prompt. More specifically, some research focuses on highlighting the item identity, which is vital for RS, in the prompt. Chu \etal~\cite{chu2023leveraging} designs a novel mask mechanism and position embedding to distinguish the items from the lingual input when fine-tuning a GLM model. Furthermore, E4SRec~\cite{li2023e4srec} proposes to use the ID embedding accompanied with a linear projection to represent the items in the prompt. RecInterpreter~\cite{yang2023large} and LLaRA~\cite{liao2023llara} share a similar idea of E4SRec, while they apply a pre-trained sequential recommender to encode the item identities. The other line is non-tunable LLM4Rec, which is mainly devoted to designing the process flow for hyperscale LLMs, such as ChatGPT and GPT-4. For instance, Chat-Rec~\cite{gao2023chat} reformulates the recommendation task into a conversational process, and thus can utilize the ChatGPT to give out proper recommendations.
Hou \etal~\cite{hou2023large} propose a combination of several types of prompts to improve the ranking performance.

Though existing works have taken an early step to adapt LLMs to recommendation, they still face several challenges, such as high inference costs and out-of-corpus. In this paper, we propose a novel method to address these two issues.

\subsection{Medication Recommendation}

Medication recommendation have been highlighted in recent years, because of their practical values. In the early stage of the research, some works aim to model the relationship between the diagnosis and prescriptions in the current visit carefully. For example, Leap~\cite{zhang2017leap} captures the mutual effects between several diagnoses and models the recommendation as a sequential decision-making process. Later, 4SDrug~\cite{tan20224sdrug} proposes to measure the similarity between symptom and medication sets for the recommendation. Furthermore, Zhang \etal~\cite{zhang2023knowledge} fabricates graph-based architecture to embed the relations between symptoms and medications via the knowledge graph and attributes. Compared with the models only using the information from the current visit, many other works target modeling the historical treatment records for better performance. RETAIN~\cite{choi2016retain} firstly develops a time-series prediction model for healthcare specially. GAMENet~\cite{shang2019gamenet} and SafeDrug~\cite{yang2021safedrug} both utilize the historical diagnosis and procedure data for medication recommendation and consider the problem of drug-drug interaction. G-Bert~\cite{shang2019pre} introduces the pre-train technique to get a better diagnosis and medication encoders for the final recommendation. Moreover, some works further intake the prescription history, which is an important reference to the recommendation at that time. For instance, MICRON~\cite{yang2021change} and COGNet~\cite{wu2022conditional} both consider copying the historical prescriptions to the current recommending drug set in a certain probability. REFINE~\cite{bhoi2023refine} directly inputs the records into a transformer encoder for modeling. However, the existing models only utilize the identities to obtain the collaborative information, while ignoring the medical semantics contained in EHR. As far as we know, we are the first to combine the LLM with medication recommendation for acquiring semantic knowledge.

%% file: 6Conclusion.tex
\section{Conclusion}

In this paper, we propose a large language model enhanced medication by distillation (LEADER). To adapt the large language model to the medication recommendation task, we first design the proper prompt templates to derive the lingual input for LLM. Then, we substitute the head layer of LLM to alleviate the out-of-corpus problem and adopt the BCE loss to fine-tune the modified LLM. However, the LLM-based model faces the challenge of high inference costs. For higher efficiency, we devise a feature-level knowledge distillation method to transfer the powerful ability of LLM to a relatively small student model. Through extensive experiments on two public datasets, we have verified that the proposed LEADER can achieve effective and efficient medication recommendation compared with existing state-of-the-art models. In terms of future work, we will consider the drug-drug interaction in LLM-based medication recommendation, which is related to the safety of prescriptions.


%% file: 7Appendix.tex
\section{Experimental Settings}


\subsection{Dataset}    \label{sec:appendix_dataset}

We adopt the common-used datasets, MIMIC-III and MIMIC-IV, in the experiments. They are from a medical database, named Medical Information Mart for Intensive Care (MIMIC)~\footnote{https://mimic.mit.edu/}. The duration of the MIMIC-III is from 2001 to 2012, while MIMIC-IV is from 2008 to 2019. Following the preprocessing of the previous works~\cite{shang2019gamenet,yang2021safedrug}, we transform the NDC code of the medications to ATC level codes to get the drug-drug interaction (DDI) graph and molecule connection graph for the implementation of the baselines. Besides, we only retain the prescriptions during the first 24 hours of a visit, as previous works~\cite{shang2019gamenet,yang2021safedrug} did. We filter out the medications that cannot be mapped to ATC codes and the visits that have void input set. At last, we split the data into train/validation/test by the ratio of 8:1:1. The statistics of the preprocessed data are shown in Table~\ref{tab:exp_dataset}.

\begin{table}[!t]
\centering
\caption{The statistics of the preprocessed datasets}
\resizebox{0.5\textwidth}{!}{
\begin{tabular}{c|cc}
\toprule[1pt]
\textbf{Item} & \textbf{MIMIC-III} & \textbf{MIMIC-IV} \\ \midrule
\# of single-visit patients & 908 & 2,877 \\
\# of multi-visit patients & 5,442 & 6,029 \\
diag. / proc. / med. space size & 1,958 / 1,430 / 112 & 1,998 / 1,001 / 125 \\
avg. / max of diag. per visit & 10.51 / 128 & 8.41 / 220 \\
avg. / max of proc. per visit & 3.84 / 50 & 2.11 / 49 \\
avg. / max of med. per visit & 11.64 / 64 & 7.02 / 72 \\ 
\bottomrule[1pt]
\end{tabular}
}
\label{tab:exp_dataset}
\vspace{-4mm}
\end{table}

\subsection{Baselines}  \label{sec:appendix_baseline}


    \noindent \textbf{Medication Recommendation Model}. We compare with many up-to-date medication recommendation models in the experiments. 
    \begin{itemize}[leftmargin=*]
        \item \textbf{RETAIN}~\cite{choi2016retain}. RETAIN designs a two-level attention model to enhance the accuracy and interpretability of clinical variable prediction. We implement it by adding the representation of diagnosis and procedure for each visit.
        \item \textbf{G-Bert}~\cite{shang2019pre}. G-Bert utilizes all the data to pre-train diagnosis and medication encoders, but needs historical medication records in fine-tuning stage.
        \item \textbf{GAMENet}~\cite{shang2019gamenet}. GAMENet adopts the memory bank to integrate global medication interaction and drug-drug interaction knowledge. For the implementation, we substitute the retrieval representation from patient history with the one from patient similarity for those single-visit patients.
        \item \textbf{SafeDrug}~\cite{yang2021safedrug}. SafeDrug utilizes drug molecule structure to encode the medications and add direct drug-drug interaction control during the training process. 
        \item \textbf{MICRON}~\cite{yang2021change}. MICRON finds that there is little distinction between the prescription in two successive visits and thus captures the change for the final recommendation. For the input, the medication set taken on the last visit is compulsory.
        \item \textbf{COGNet}~\cite{wu2022conditional}. COGNet copies the ever-prescribed drugs to the current visit, so the previous medication records are necessary.
        \item \textbf{REFINE}~\cite{bhoi2023refine}. REFINE proposes to model the severity of the drug-drug interaction and the fine-grained medication dosage. For a fair comparison, we implement it by inputting the diagnosis rather than lab test responses to the model. Since REFINE also takes the historical medication records as input, it cannot infer for single-visit patients.
    \end{itemize}

    \noindent \textbf{LLM-based Recommendation Model}. Some research studies have proposed to utilize the large language model to complete the recommendation task. To further verify the effectiveness of our LEADER, we compare with them in the experiments. 
    \begin{itemize}[leftmargin=*]
        \item \textbf{TALLRec}~\cite{bao2023tallrec}. TALLRec is one of the pioneering works to adapt the LLM to the recommendation task. It proposed constructing the recommendation as a text generation task and then instruction tuning the open-sourced LLM. In the experiments, we adopt the same prompts as our LEADER to motivate the LLM to complete medication recommendation. 
        \item \textbf{BIGRec}~\cite{bao2023bi}. To alleviate the out-of-corpus problem for LLM, BIGRec proposed a two-step framework to ground the actual items. For the implementation, we achieve the first step same as the TALLRec. Then, we calculate the LLM embedding of textual patient prompts and medication names to calculate the recommending probability of each drug.
        \item \textbf{E4SRec}~\cite{li2023e4srec}. E4SRec proposed to integrate the pre-trained collaborative embeddings into the LLM for the sequential recommendation. In our implementation, we insert the medication, diagnosis and procedure embeddings of the pre-trained REFINE~\cite{bhoi2023refine} into the corresponding positions of the patient prompt.
    \end{itemize}

\begin{table*}[!t]
\centering
\tabcolsep=0.05cm   
\caption{Performance (PRAUC) comparison of the proposed LEADER based on LLaMA-7B, Qwen-7B and Qwen-1.8B.}
\label{tab:senstivity}
\begin{tabular}{cc|ccc|ccc}
\toprule[1pt]
\multirow{2}{*}{\textbf{LLM}}       & \multirow{2}{*}{\textbf{Model}} & \multicolumn{3}{c|}{\textbf{MIMIC-III}}       & \multicolumn{3}{c}{\textbf{MIMIC-IV}}         \\ 
\cmidrule{3-8} 
&   & \textbf{Overall} & \textbf{Multi-visit} & \textbf{Single-visit} & \textbf{Overall} & \textbf{Multi-visit} & \textbf{Single-visit} \\ 
\midrule
\multirow{2}{*}{LLaMA-7B}  
& LEADER(T)  & 0.7816 $\pm$ 0.0015 & 0.7854 $\pm$ 0.0015 & 0.7590 $\pm$ 0.0046 & 0.7120 $\pm$0.0024 & 0.7238 $\pm$ 0.0031 & 0.6881 $\pm$ 0.0039  \\
& LEADER(S)  & 0.7795 $\pm$ 0.0025 & 0.7830 $\pm$ 0.0019 & 0.7631 $\pm$ 0.0056 & 0.7020 $\pm$ 0.0022 & 0.6994 $\pm$ 0.0037 & 0.7033 $\pm$ 0.0041             \\ 
\midrule
\multirow{2}{*}{Qwen-7B}   
& LEADER(T) & 0.7822 $\pm$ 0.0021 & 0.7884 $\pm$ 0.0015 & 0.7584 $\pm$ 0.0050 & 0.7145 $\pm$ 0.0019 & 0.7321 $\pm$ 0.0035 & 0.6896 $\pm$ 0.0033 \\
& LEADER(S) & 0.7781 $\pm$ 0.0017 & 0.7849 $\pm$ 0.0012 & 0.7563 $\pm$ 0.0057 & 0.6999 $\pm$ 0.0023 & 0.7061 $\pm$ 0.0045 & 0.6975 $\pm$ 0.0043 \\ 
\midrule
\multirow{2}{*}{Qwen-1.8B} 
& LEADER(T)  & 0.7662 $\pm$ 0.0013 & 0.7735 $\pm$ 0.0017 & 0.7403 $\pm$ 0.0056 & 0.7026 $\pm$ 0.0027 & 0.7140 $\pm$ 0.0027 & 0.6831 $\pm$ 0.0063 \\
& LEADER(S) & 0.7740 $\pm$ 0.0015 & 0.7783 $\pm$ 0.0015 & 0.7593 $\pm$ 0.0042 & 0.6950 $\pm$ 0.0024 & 0.6947 $\pm$ 0.0042 & 0.6943 $\pm$ 0.0034 \\ 
\bottomrule[1pt]
\end{tabular}
\end{table*}

\subsection{Implementation Details} \label{sec:appendix_implement}

    The hardware used in the experiments is an Intel Xeon Gold 6133 platform with Tesla V100 32G GPUs, while the software basis includes Python 3.9.5 and PyTorch 1.12.0. For our LEADER(T) and the compared LLM-based baselines, we all adopt the LLaMA-7B~\cite{touvron2023llama} as the foundation model. For the sensitive analysis in the later Section~\ref{sec:appendix_llm}, we additionally adopt \ie Qwen-7B~\cite{qwen} in the experiments.
    During the fine-tuning, the LoRA~\cite{hu2021lora} layers are accompanied by the layers identified as ``q\_proj'', ``k\_proj'', ``v\_proj'', ``o\_proj'', ``down\_proj'', ``up\_proj'' and ``gate\_proj'' in LLM. Other configurations include LoRA rank of $8$, batch size of $32$, learning rate of $2e-4$ and maximum input length of $2,048$. Due to different data scales, the maximum training steps are set to 3,000 and 4,000 for MIMIC-III and MIMIC-IV, respectively. In terms of the distillation for the student model LEADER(S), we set the dimension $d_e$ and $d_t$ to $64$, the number of transformer layers of all encoders $\mathcal{E}$ to $1$ and $\tau$ to $1$. We adopt Adam optimizer~\cite{kingma2014adam} and set the learning rate to $5e-4$. The batch size is fixed at $4$ for MIMIC-III and $16$ for MIMIC-IV. The best hyper-parameters are chosen by the PRAUC metric on the validation set. Specifically, $\alpha$ is tuned from $0.1$ to $0.9$, while $\beta$ is searched from $\{0.1, 0.05, 0.01, 0.005, 0.001\}$.

\section{Supplymentary Results}


\subsection{Sentivity Analysis} \label{sec:appendix_llm}

    To probe the effect of the type and size of LLM for our LEADER, we conduct the experiments on the proposed LEADER under LLaMA-7B~\cite{touvron2023llama}, Qwen-7B~\cite{qwen} and Qwen-1.8B~\cite{qwen}. The results on the two datasets are shown in Table~\ref{tab:senstivity}. 
    The performance comparison indicates that the LEADER(T) and LEADER(S) both show almost the same tendency with the LLaMA-7B and Qwen-7B, which illustrates that the designed method is insensitive to the type of LLMs under the same parameter scales. More detailed, Qwen-based LEADER(T) shows a slight superiority over LLaMA-based LEADER(T), which is caused by better recommending performance in the multi-visit group. The reason may lie in that the Qwen-7B has a longer input length limitation than LLaMA-7B, and the prompt for multi-visit patients is rather complex. 
    Then, we observe that the performance of Qwen-1.8B is lower than that of Qwen-7B for both the teacher and student models. This finding highlights that larger-scale models deliver better recommendations due to their enhanced semantic understanding capabilities. Moreover, we notice that the performance drop for LEADER(S) is less pronounced compared to LEADER(T) when the model size decreases from 7B to 1.8B. This suggests that the effect of LLM scale is more significant for the teacher model than the student model. 

%% file: main.bbl
\begin{thebibliography}{10}
\providecommand{\url}[1]{#1}
\csname url@samestyle\endcsname
\providecommand{\newblock}{\relax}
\providecommand{\bibinfo}[2]{#2}
\providecommand{\BIBentrySTDinterwordspacing}{\spaceskip=0pt\relax}
\providecommand{\BIBentryALTinterwordstretchfactor}{4}
\providecommand{\BIBentryALTinterwordspacing}{\spaceskip=\fontdimen2\font plus
\BIBentryALTinterwordstretchfactor\fontdimen3\font minus \fontdimen4\font\relax}
\providecommand{\BIBforeignlanguage}[2]{{%
\expandafter\ifx\csname l@#1\endcsname\relax
\typeout{** WARNING: IEEEtran.bst: No hyphenation pattern has been}%
\typeout{** loaded for the language `#1'. Using the pattern for}%
\typeout{** the default language instead.}%
\else
\language=\csname l@#1\endcsname
\fi
#2}}
\providecommand{\BIBdecl}{\relax}
\BIBdecl

\bibitem{choi2016retain}
E.~Choi, M.~T. Bahadori, J.~Sun, J.~Kulas, A.~Schuetz, and W.~Stewart, ``Retain: An interpretable predictive model for healthcare using reverse time attention mechanism,'' \emph{Advances in neural information processing systems}, vol.~29, 2016.

\bibitem{shang2019pre}
J.~Shang, T.~Ma, C.~Xiao, and J.~Sun, ``Pre-training of graph augmented transformers for medication recommendation,'' \emph{arXiv preprint arXiv:1906.00346}, 2019.

\bibitem{shang2019gamenet}
J.~Shang, C.~Xiao, T.~Ma, H.~Li, and J.~Sun, ``Gamenet: Graph augmented memory networks for recommending medication combination,'' in \emph{proceedings of the AAAI Conference on Artificial Intelligence}, vol.~33, no.~01, 2019, pp. 1126--1133.

\bibitem{yang2021safedrug}
C.~Yang, C.~Xiao, F.~Ma, L.~Glass, and J.~Sun, ``Safedrug: Dual molecular graph encoders for recommending effective and safe drug combinations,'' \emph{arXiv preprint arXiv:2105.02711}, 2021.

\bibitem{yang2021change}
C.~Yang, C.~Xiao, L.~Glass, and J.~Sun, ``Change matters: Medication change prediction with recurrent residual networks,'' \emph{arXiv preprint arXiv:2105.01876}, 2021.

\bibitem{wu2022conditional}
R.~Wu, Z.~Qiu, J.~Jiang, G.~Qi, and X.~Wu, ``Conditional generation net for medication recommendation,'' in \emph{Proceedings of the ACM Web Conference 2022}, 2022, pp. 935--945.

\bibitem{bhoi2023refine}
S.~Bhoi, M.-L. Lee, W.~Hsu, and N.~C. Tan, ``Refine: A fine-grained medication recommendation system using deep learning and personalized drug interaction modeling,'' in \emph{Thirty-seventh Conference on Neural Information Processing Systems}, 2023.

\bibitem{rahmawati2020physician}
I.~Rahmawati and V.~I.~D. Prastika, ``Physician knowledge and responsibility of prescription policy,'' \emph{Jurnal Administrasi Kesehatan Indonesia Volume}, vol.~8, no.~1, 2020.

\bibitem{ali2023deep}
Z.~Ali, Y.~Huang, I.~Ullah, J.~Feng, C.~Deng, N.~Thierry, A.~Khan, A.~U. Jan, X.~Shen, W.~Rui \emph{et~al.}, ``Deep learning for medication recommendation: a systematic survey,'' \emph{Data Intelligence}, vol.~5, no.~2, pp. 303--354, 2023.

\bibitem{li2023chatdoctor}
Y.~Li, Z.~Li, K.~Zhang, R.~Dan, and Y.~Zhang, ``Chatdoctor: A medical chat model fine-tuned on llama model using medical domain knowledge,'' \emph{arXiv e-prints}, pp. arXiv--2303, 2023.

\bibitem{zhao2023survey}
W.~X. Zhao, K.~Zhou, J.~Li, T.~Tang, X.~Wang, Y.~Hou, Y.~Min, B.~Zhang, J.~Zhang, Z.~Dong \emph{et~al.}, ``A survey of large language models,'' \emph{arXiv preprint arXiv:2303.18223}, 2023.

\bibitem{wei2022chain}
J.~Wei, X.~Wang, D.~Schuurmans, M.~Bosma, F.~Xia, E.~Chi, Q.~V. Le, D.~Zhou \emph{et~al.}, ``Chain-of-thought prompting elicits reasoning in large language models,'' \emph{Advances in Neural Information Processing Systems}, vol.~35, pp. 24\,824--24\,837, 2022.

\bibitem{borisov2022language}
V.~Borisov, K.~Sessler, T.~Leemann, M.~Pawelczyk, and G.~Kasneci, ``Language models are realistic tabular data generators,'' in \emph{The Eleventh International Conference on Learning Representations}, 2022.

\bibitem{chen2023large}
J.~Chen, Z.~Liu, X.~Huang, C.~Wu, Q.~Liu, G.~Jiang, Y.~Pu, Y.~Lei, X.~Chen, X.~Wang \emph{et~al.}, ``When large language models meet personalization: Perspectives of challenges and opportunities,'' \emph{arXiv preprint arXiv:2307.16376}, 2023.

\bibitem{wu2023survey}
L.~Wu, Z.~Zheng, Z.~Qiu, H.~Wang, H.~Gu, T.~Shen, C.~Qin, C.~Zhu, H.~Zhu, Q.~Liu \emph{et~al.}, ``A survey on large language models for recommendation,'' \emph{arXiv preprint arXiv:2305.19860}, 2023.

\bibitem{yang2023large}
Z.~Yang, J.~Wu, Y.~Luo, J.~Zhang, Y.~Yuan, A.~Zhang, X.~Wang, and X.~He, ``Large language model can interpret latent space of sequential recommender,'' \emph{arXiv preprint arXiv:2310.20487}, 2023.

\bibitem{lin2023multi}
X.~Lin, W.~Wang, Y.~Li, F.~Feng, S.-K. Ng, and T.-S. Chua, ``A multi-facet paradigm to bridge large language model and recommendation,'' \emph{arXiv preprint arXiv:2310.06491}, 2023.

\bibitem{zheng2023generative}
Z.~Zheng, Z.~Qiu, X.~Hu, L.~Wu, H.~Zhu, and H.~Xiong, ``Generative job recommendations with large language model,'' \emph{arXiv preprint arXiv:2307.02157}, 2023.

\bibitem{zhou2023opportunities}
Y.~Zhou, X.~Lin, X.~Zhang, M.~Wang, G.~Jiang, H.~Lu, Y.~Wu, K.~Zhang, Z.~Yang, K.~Wang \emph{et~al.}, ``On the opportunities of green computing: A survey,'' \emph{arXiv preprint arXiv:2311.00447}, 2023.

\bibitem{gruendner2019ketos}
J.~Gruendner, T.~Schwachhofer, P.~Sippl, N.~Wolf, M.~Erpenbeck, C.~Gulden, L.~A. Kapsner, J.~Zierk, S.~Mate, M.~St{\"u}rzl \emph{et~al.}, ``Ketos: Clinical decision support and machine learning as a service--a training and deployment platform based on docker, omop-cdm, and fhir web services,'' \emph{PloS one}, vol.~14, no.~10, p. e0223010, 2019.

\bibitem{zeng2022glm}
A.~Zeng, X.~Liu, Z.~Du, Z.~Wang, H.~Lai, M.~Ding, Z.~Yang, Y.~Xu, W.~Zheng, X.~Xia \emph{et~al.}, ``Glm-130b: An open bilingual pre-trained model,'' in \emph{The Eleventh International Conference on Learning Representations}, 2022.

\bibitem{touvron2023llama}
H.~Touvron, T.~Lavril, G.~Izacard, X.~Martinet, M.-A. Lachaux, T.~Lacroix, B.~Rozi{\`e}re, N.~Goyal, E.~Hambro, F.~Azhar, A.~Rodriguez, A.~Joulin, E.~Grave, and G.~Lample, ``Llama: Open and efficient foundation language models,'' \emph{arXiv preprint arXiv:2302.13971}, 2023.

\bibitem{openai2023gpt4}
OpenAI, ``Gpt-4 technical report,'' \emph{arXiv preprint arXiv:2303.08774}, 2023.

\bibitem{bao2023bi}
K.~Bao, J.~Zhang, W.~Wang, Y.~Zhang, Z.~Yang, Y.~Luo, F.~Feng, X.~He, and Q.~Tian, ``A bi-step grounding paradigm for large language models in recommendation systems,'' \emph{arXiv preprint arXiv:2308.08434}, 2023.

\bibitem{geng2022recommendation}
S.~Geng, S.~Liu, Z.~Fu, Y.~Ge, and Y.~Zhang, ``Recommendation as language processing (rlp): A unified pretrain, personalized prompt \& predict paradigm (p5),'' in \emph{Proceedings of the 16th ACM Conference on Recommender Systems}, 2022, pp. 299--315.

\bibitem{bao2023tallrec}
K.~Bao, J.~Zhang, Y.~Zhang, W.~Wang, F.~Feng, and X.~He, ``Tallrec: An effective and efficient tuning framework to align large language model with recommendation,'' \emph{arXiv preprint arXiv:2305.00447}, 2023.

\bibitem{wang2023huatuo}
H.~Wang, C.~Liu, N.~Xi, Z.~Qiang, S.~Zhao, B.~Qin, and T.~Liu, ``Huatuo: Tuning llama model with chinese medical knowledge,'' 2023.

\bibitem{hu2021lora}
E.~J. Hu, P.~Wallis, Z.~Allen-Zhu, Y.~Li, S.~Wang, L.~Wang, W.~Chen \emph{et~al.}, ``Lora: Low-rank adaptation of large language models,'' in \emph{International Conference on Learning Representations}, 2021.

\bibitem{gou2021knowledge}
J.~Gou, B.~Yu, S.~J. Maybank, and D.~Tao, ``Knowledge distillation: A survey,'' \emph{International Journal of Computer Vision}, vol. 129, pp. 1789--1819, 2021.

\bibitem{tirumala2022memorization}
K.~Tirumala, A.~Markosyan, L.~Zettlemoyer, and A.~Aghajanyan, ``Memorization without overfitting: Analyzing the training dynamics of large language models,'' \emph{Advances in Neural Information Processing Systems}, vol.~35, pp. 38\,274--38\,290, 2022.

\bibitem{biderman2023emergent}
S.~Biderman, U.~S. Prashanth, L.~Sutawika, H.~Schoelkopf, Q.~Anthony, S.~Purohit, and E.~Raf, ``Emergent and predictable memorization in large language models,'' \emph{arXiv preprint arXiv:2304.11158}, 2023.

\bibitem{li2022blip}
J.~Li, D.~Li, C.~Xiong, and S.~Hoi, ``Blip: Bootstrapping language-image pre-training for unified vision-language understanding and generation,'' in \emph{International Conference on Machine Learning}.\hskip 1em plus 0.5em minus 0.4em\relax PMLR, 2022, pp. 12\,888--12\,900.

\bibitem{radford2021learning}
A.~Radford, J.~W. Kim, C.~Hallacy, A.~Ramesh, G.~Goh, S.~Agarwal, G.~Sastry, A.~Askell, P.~Mishkin, J.~Clark \emph{et~al.}, ``Learning transferable visual models from natural language supervision,'' in \emph{International conference on machine learning}.\hskip 1em plus 0.5em minus 0.4em\relax PMLR, 2021, pp. 8748--8763.

\bibitem{chen2020simple}
T.~Chen, S.~Kornblith, M.~Norouzi, and G.~Hinton, ``A simple framework for contrastive learning of visual representations,'' in \emph{International conference on machine learning}.\hskip 1em plus 0.5em minus 0.4em\relax PMLR, 2020, pp. 1597--1607.

\bibitem{li2023e4srec}
X.~Li, C.~Chen, X.~Zhao, Y.~Zhang, and C.~Xing, ``E4srec: An elegant effective efficient extensible solution of large language models for sequential recommendation,'' \emph{arXiv preprint arXiv:2312.02443}, 2023.

\bibitem{hinton2015distilling}
G.~Hinton, O.~Vinyals, and J.~Dean, ``Distilling the knowledge in a neural network,'' \emph{arXiv preprint arXiv:1503.02531}, 2015.

\bibitem{li2023large}
L.~Li, Y.~Zhang, D.~Liu, and L.~Chen, ``Large language models for generative recommendation: A survey and visionary discussions,'' \emph{arXiv preprint arXiv:2309.01157}, 2023.

\bibitem{fan2023recommender}
W.~Fan, Z.~Zhao, J.~Li, Y.~Liu, X.~Mei, Y.~Wang, J.~Tang, and Q.~Li, ``Recommender systems in the era of large language models (llms),'' \emph{arXiv preprint arXiv:2307.02046}, 2023.

\bibitem{raffel2020exploring}
C.~Raffel, N.~Shazeer, A.~Roberts, K.~Lee, S.~Narang, M.~Matena, Y.~Zhou, W.~Li, and P.~J. Liu, ``Exploring the limits of transfer learning with a unified text-to-text transformer,'' \emph{The Journal of Machine Learning Research}, vol.~21, no.~1, pp. 5485--5551, 2020.

\bibitem{liu2023moelora}
Q.~Liu, X.~Wu, X.~Zhao, Y.~Zhu, D.~Xu, F.~Tian, and Y.~Zheng, ``When moe meets llms: Parameter efficient fine-tuning for multi-task medical applications,'' in \emph{Proceedings of the 47th International ACM SIGIR Conference on Research and Development in Information Retrieval}, 2024, pp. 1104--1114.

\bibitem{zhang2023recommendation}
J.~Zhang, R.~Xie, Y.~Hou, W.~X. Zhao, L.~Lin, and J.-R. Wen, ``Recommendation as instruction following: A large language model empowered recommendation approach,'' \emph{arXiv preprint arXiv:2305.07001}, 2023.

\bibitem{chen2023palr}
Z.~Chen, ``Palr: Personalization aware llms for recommendation,'' \emph{arXiv preprint arXiv:2305.07622}, 2023.

\bibitem{chu2023leveraging}
Z.~Chu, H.~Hao, X.~Ouyang, S.~Wang, Y.~Wang, Y.~Shen, J.~Gu, Q.~Cui, L.~Li, S.~Xue \emph{et~al.}, ``Leveraging large language models for pre-trained recommender systems,'' \emph{arXiv preprint arXiv:2308.10837}, 2023.

\bibitem{liao2023llara}
J.~Liao, S.~Li, Z.~Yang, J.~Wu, Y.~Yuan, X.~Wang, and X.~He, ``Llara: Aligning large language models with sequential recommenders,'' \emph{arXiv preprint arXiv:2312.02445}, 2023.

\bibitem{gao2023chat}
Y.~Gao, T.~Sheng, Y.~Xiang, Y.~Xiong, H.~Wang, and J.~Zhang, ``Chat-rec: Towards interactive and explainable llms-augmented recommender system,'' \emph{arXiv preprint arXiv:2303.14524}, 2023.

\bibitem{hou2023large}
Y.~Hou, J.~Zhang, Z.~Lin, H.~Lu, R.~Xie, J.~McAuley, and W.~X. Zhao, ``Large language models are zero-shot rankers for recommender systems,'' \emph{arXiv preprint arXiv:2305.08845}, 2023.

\bibitem{zhang2017leap}
Y.~Zhang, R.~Chen, J.~Tang, W.~F. Stewart, and J.~Sun, ``Leap: learning to prescribe effective and safe treatment combinations for multimorbidity,'' in \emph{proceedings of the 23rd ACM SIGKDD international conference on knowledge Discovery and data Mining}, 2017, pp. 1315--1324.

\bibitem{tan20224sdrug}
Y.~Tan, C.~Kong, L.~Yu, P.~Li, C.~Chen, X.~Zheng, V.~S. Hertzberg, and C.~Yang, ``4sdrug: Symptom-based set-to-set small and safe drug recommendation,'' in \emph{Proceedings of the 28th ACM SIGKDD Conference on Knowledge Discovery and Data Mining}, 2022, pp. 3970--3980.

\bibitem{zhang2023knowledge}
Y.~Zhang, X.~Wu, Q.~Fang, S.~Qian, and C.~Xu, ``Knowledge-enhanced attributed multi-task learning for medicine recommendation,'' \emph{ACM Transactions on Information Systems}, vol.~41, no.~1, pp. 1--24, 2023.

\bibitem{qwen}
J.~Bai, S.~Bai, Y.~Chu, Z.~Cui, K.~Dang, X.~Deng, Y.~Fan, W.~Ge, Y.~Han, F.~Huang, B.~Hui, L.~Ji, M.~Li, J.~Lin, R.~Lin, D.~Liu, G.~Liu, C.~Lu, K.~Lu, J.~Ma, R.~Men, X.~Ren, X.~Ren, C.~Tan, S.~Tan, J.~Tu, P.~Wang, S.~Wang, W.~Wang, S.~Wu, B.~Xu, J.~Xu, A.~Yang, H.~Yang, J.~Yang, S.~Yang, Y.~Yao, B.~Yu, H.~Yuan, Z.~Yuan, J.~Zhang, X.~Zhang, Y.~Zhang, Z.~Zhang, C.~Zhou, J.~Zhou, X.~Zhou, and T.~Zhu, ``Qwen technical report,'' \emph{arXiv preprint arXiv:2309.16609}, 2023.

\bibitem{kingma2014adam}
D.~P. Kingma and J.~Ba, ``Adam: A method for stochastic optimization,'' \emph{arXiv preprint arXiv:1412.6980}, 2014.

\end{thebibliography}
